 \documentclass[aip,pop,reprint]{revtex4-1}
 \usepackage{dcolumn}
 \usepackage{bm}
 \usepackage{amsmath}
 
\newcommand{\arctanh}{\textrm{arctanh}}

\newcommand{\sech}{\textrm{sech}}

\newcommand{\vc}{\mathbf}
\usepackage{graphicx}
\usepackage{color}

\begin{document}


\title{Collisionless Kinetic Theory of Oblique Tearing Instabilities}


\author{S.\ D.\ Baalrud}
\affiliation{Department of Physics and Astronomy, University of Iowa, Iowa City, Iowa 52242, USA}

\author{A.\ Bhattacharjee}
\affiliation{Department of Astrophysical Sciences, Princeton University, Princeton, New Jersey 08544, USA}

\author{W.\ Daughton}
\affiliation{Los Alamos National Laboratory, Los Alamos, New Mexico 87545, USA}


\date{\today}

\begin{abstract}


The linear dispersion relation for collisionless kinetic tearing instabilities is calculated for a Harris equilibrium.
In contrast to the conventional 2D geometry, which considers only modes at the center of the current sheet, modes can span the current sheet in 3D. 
Modes at each resonant surface have a unique angle ($\theta$) with respect to the guide field direction. 
Both kinetic simulations and numerical eigenmode solutions of the linearized Vlasov-Maxwell equations have recently revealed that standard analytic theories vastly overestimate the growth rate of oblique modes ($\theta \neq 0$). 
We find that this stabilization is associated with the density-gradient-driven diamagnetic drift. 
The analytic theories miss this drift stabilization because the inner tearing layer broadens at oblique angles sufficiently far that the assumption of scale separation between the inner and outer regions of boundary-layer theory breaks down. 
The dispersion relation obtained by numerically solving a single second order differential equation is found to approximately capture the drift stabilization predicted by solutions of the full integro-differential eigenvalue problem. 
A simple analytic estimate for the stability criterion is provided. 


\end{abstract}




\maketitle


\section{Introduction}


Tearing instabilities are magnetic reconnection events that occur in transition layers where the frozen flux condition of ideal magnetohydrodynamics (MHD) is violated.~\cite{furt:63} 
In tokamaks, they contribute to the basic magnetic field topology,~\cite{hegn:93} transport,\cite{call:77,biew:03,doer:11,gutt:11} rotation,~\cite{cole:15} and may trigger disruptive events.~\cite{butt:00} 
In space, they are responsible for initiating large-scale magnetic reconnection in Earth's magnetosphere,~\cite{daug:05} as well as flares in the solar corona.~\cite{inne:15} 
If the plasma is sufficiently dense or cool, tearing instabilities are well described by resistive,~\cite{furt:63,lour:07,bhat:09,cass:09,lour:12,huan:12} or visco-resistive~\cite{comi:16} MHD.
At lower densities or higher temperatures, two-fluid or Hall MHD effects modify the resistive reconnection rate.~\cite{fitz:04,mirn:04,baal:11,huan:11,jara:16} 
At even lower densities or higher temperatures, resistivity plays no role and reconnection is determined entirely by collisionless two-fluid or kinetic effects.~\cite{drak:77,gale:86,daug:09,roge:07,numa:11}  
Space plasmas are frequently in this latter collisionless regime, and fusion plasmas can also reach this regime.~\cite{wang:90,ques:81,ques:81b,ques:85a} 

\begin{figure}
\includegraphics[width=6.5cm]{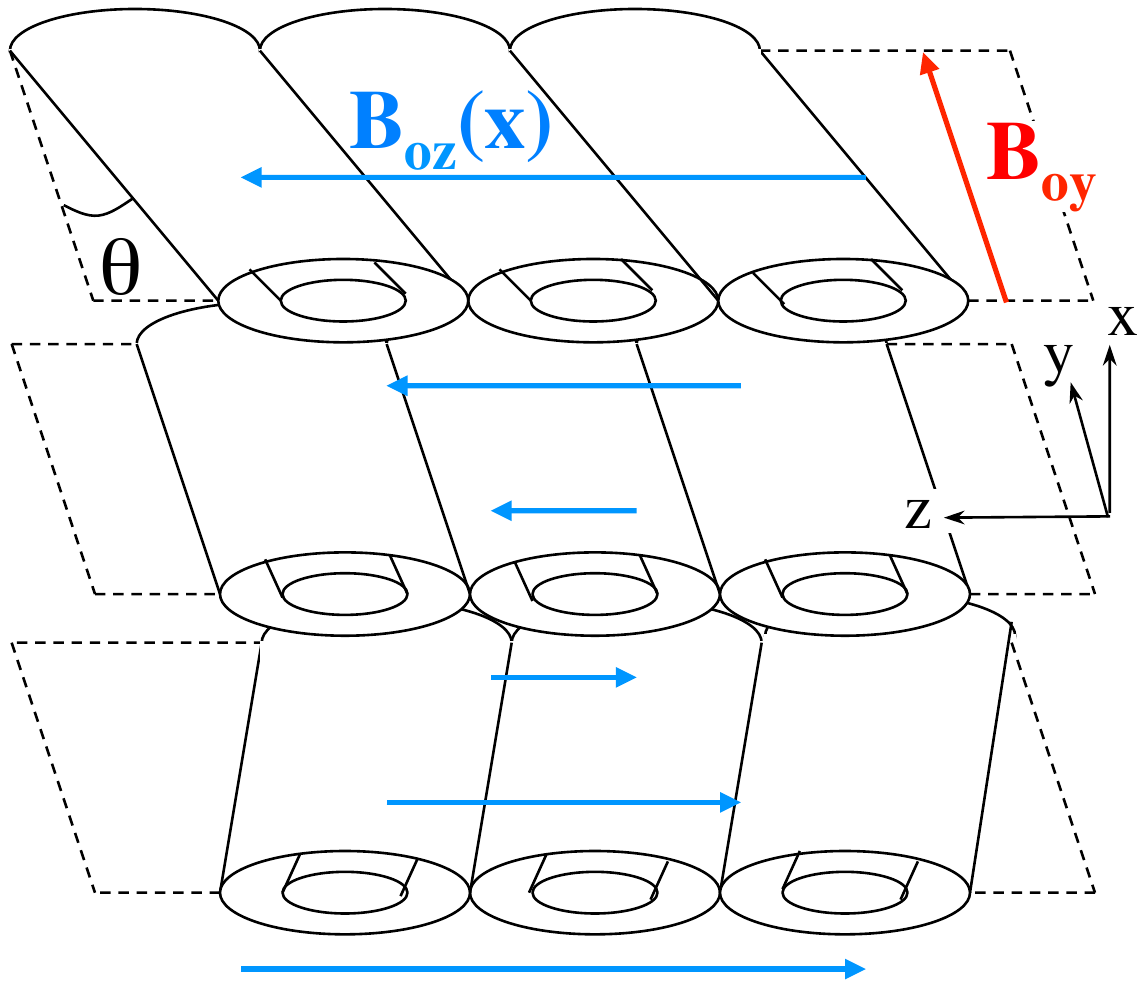}
\caption{Illustration of chains of flux tubes on three resonant surfaces. Flux tubes are aligned along the magnetic field, and $\theta$ is the angle between this and the guide field direction ($\hat{y}$). }
\label{fg:tube_draw}
\end{figure}

In this paper, we revisit the linear kinetic dispersion relation for collisionless tearing instability of a Harris current sheet.~\cite{harr:62} 
Recent particle-in-cell (PIC) simulations, together with numerical solutions of the linearized Vlasov equation,~\cite{daug:11} have revealed the surprising result that the spectrum of oblique modes is much more narrow than standard analytic theories predict.\cite{drak:77,gale:86} 
Here, $\theta = \arctan (k_y/k_z)$ is the angle of obliquity and $\vc{k}$ the wavevector; see Fig.~\ref{fg:tube_draw}. 
The resonant surface locations $x_s = \lambda \textrm{arctanh}(\mu)$, where $\mu \equiv \tan \theta B_{oy}/\bar{B}_{oz}$, locate the chain of magnetic islands associated with angle $\theta$. 
This narrower spectrum has significant consequences for the subsequent nonlinear reconnection dynamics. 
Flux ropes on adjacent resonant surfaces can overlap, which leads to stochastic magnetic field generation and even turbulence.~\cite{fuji:12,kari:13} 
The spectrum of unstable tearing modes determines the plasma volume that is ultimately susceptible to this form of turbulence. 
It also has consequences for theories of particle acceleration by contracting magnetic islands, which relies on a broad spectrum of unstable modes (i.e., volume filling of magnetic islands).~\cite{drak:06,chen:08} 
Resonant surfaces unstable to linear tearing modes also become the sites of electron layers in the nonlinear regime,\cite{liu:13} which can lead to multiple electron layers in domains long enough to support multiple resonant surfaces. Understanding the spectrum of linear tearing modes is critical. 

This work focuses on understanding the stabilization of oblique modes.
There are many theories of linear collisionless tearing modes in the literature,\cite{drak:77,gale:86,haze:75,haze:78a,haze:78b,maha:78,drak:78,lee:80,buss:78,copp:79,hosh:87,maha:79} but none of them captures this effect. We focus on comparing with Drake and Lee,\cite{drak:77} and Galeev \emph{et al},~\cite{gale:86} as these are the most commonly used. 
In a kinetic plasma, the linear response is determined by orbit integrals, leading to an eigenvalue problem with complex integro-differential operators. 
All of the theories make assumptions to reduce the complexity of these equations, including various combinations of a small gyroradius expansion, the constant-$\psi$ approximation, neglect of electrostatic terms, etc. Each of these theories also assumes a scale separation between an ``inner'' region where reconnection occurs and an ``outer'' (ideal) region. 

To identify the approximation responsible for missing the effect, we apply a series of three successive approximations and discuss the influence of each on the growth rate.
The most general description is a normal mode analysis of the Vlasov-Maxwell integro-differential (VMID) equations that includes the full orbit integrals. 
This full system of equations has been solved numerically for the Harris sheet geometry using both Hermite~\cite{daug:99} and finite-element expansions~\cite{daug:03,daug:05} of the eigenfunction. In this paper, the results from the finite-element code~\cite{daug:03} are used to evaluate and refine simplified theories.  
As a first step, the full system of equations is simplified by taking the small gyroradius limit. 
The next step exploits a scale separation that allows one to eliminate electrostatic terms from the system of equations. 
We find that each set of equations captures the narrower mode spectrum, in qualitative agreement with the previous kinetic simulations and the full numerical solutions of the tearing eigenmode equations. Quantitative differences do arise at each level of approximation. 

We find that the physical process responsible for the stabilization of oblique modes is the uniform diamagnetic drift in the Harris equilibrium.  
The drift broadens the inner layer of oblique modes. 
At large angles, the inner region becomes broad enough that the scale separation that boundary layer theories are predicated on breaks down. 
In this regime, both kinetic simulations and VMID calculations indicate that tearing modes are damped. 
In this work, we demonstrate that this stabilization effect can be described from the numerical solution of a single second order differential equation. 
A comparison with previous boundary layer theories is provided, which details asymptotic limits and where they agree and disagree with the more complete theory.~\cite{drak:77,gale:86,lee:80,maha:79} 
A simple analytic formula for the cutoff angle at which stabilization occurs is provided. 

It is noteworthy that previous work on other models, or for other equilibria, have not seen the same stabilization of oblique modes. For example, modes were found to span the current sheet in resistive MHD.~\cite{baal:12} Simulations and collisionless kinetic theory, similar to what are discussed here, were found to agree well for a force-free equilibrium.~\cite{liu:13} 
Similarly, two-fluid theory was found to accurately capture the dispersion of oblique modes in this case.~\cite{akca:16} These results are consistent with what is presented here because the force-free equilibrium does not have a diamagnetic drift. We also note that Cowley \emph{et al}\cite{cowl:86} have shown that temperature gradients can stabilize tearing modes, but the density gradient did not cause stabilization in that analysis 
and it produced only logarithmic corrections to the growth rates calculated without a density gradient~\cite{wang:90}. 
There is no temperature gradient in the Harris equilibrium.

The remainder of this paper is organized as follows. The three sets of equations obtained form successive approximations of the Vlasov-Maxwell system are derived in Sec.~\ref{sec:be}. Numerical solutions are provided in Sec.~\ref{sec:numsol}. Section~\ref{sec:compare} gives a comparison of these results with previous theories. Section~\ref{sec:const_psi} shows an analysis of how the boundary layer theories break down at oblique angles and provides a simple analytic estimate for the cuttoff angle.  

\section{Basic Equations \label{sec:be}}

\subsection{Vlasov-Maxwell\label{sec:vlasov}}

This section describes the Vlasov-Maxwell integro-differential equations in which the full orbit integrals are solved numerically. 
We will refer to this as the VMID model. 
It starts from the linearized Vlasov equation, 
\begin{equation}
\frac{d \delta f_s}{dt} = - \frac{q_s}{m_s} \biggl( \delta \vc{E} + \frac{\vc{v} \times \delta \vc{B}}{c} \biggr) \cdot \frac{\partial f_{so}}{\partial \vc{v}} \label{eq:vlasov}
\end{equation}
where $d/dt = \partial/\partial t + \vc{v} \cdot \nabla + (q_s/m_s)(\vc{E}_o + \vc{v} \times \vc{B}_o/c) \cdot \nabla_\vc{v}$ is the convective derivative. We will later focus on the Harris equilibrium, but in this section require only that the equilibrium magnetic field can be written in the form $\vc{B}_o = B_{yo} (x) \hat{y} + B_{zo} (x) \hat{z}$ and that the lowest-order distribution functions are Maxwellians 
\begin{equation}
f_{so} = \frac{n_s (x)}{\pi^{3/2} v_{Ts}^3} \exp \biggl[- \frac{(\vc{v} - \vc{U}_s)^2}{v_{Ts}^2} \biggr]  \label{eq:maxwellian}
\end{equation}
where the flow velocity is $\vc{U}_s = U_{s,y} (x) \hat{y} + U_{s,z} (x) \hat{z}$. We write $\delta \vc{E}$ and $\delta \vc{B}$ in terms of scalar and vector potentials 
\begin{equation}
\delta \vc{E} = - \nabla \delta \phi - \frac{1}{c} \frac{\partial \delta \vc{A}}{\partial t} \ \ \ \textrm{and} \ \ \ \delta \vc{B} = \nabla \times \delta \vc{A}  \label{eq:vectorps}
\end{equation}
and apply a normal mode analysis of the form
\begin{align*}
\delta \phi = \delta \bar{\phi} (x) \exp (- i \omega t + i k_y y + i k_z z) , \\ 
\delta \vc{A} = \delta \bar{\vc{A}}(x) \exp (- i \omega t + i k_y y + i k_z z)  .
\end{align*}

Inserting Eq.~(\ref{eq:maxwellian}) into (\ref{eq:vlasov}) and integrating along the characteristics, which are the unperturbed particle orbits, provides\cite{daug:99}
\begin{equation}
\delta f_s = - \frac{q_s f_{so}}{T_s} \biggl[ \delta \phi - \frac{\vc{U}_s \cdot \delta \vc{A}}{c} + i (\omega - \vc{k} \cdot \vc{U}_s) S \biggr]  \label{eq:dfs1}
\end{equation}
where
\begin{equation}
S = \int_0^\infty d\tau ( \delta \bar{\phi} - \vc{v}^\prime \cdot \delta \bar{\vc{A}}/c ) e^{i \omega \tau - i \vc{k} \cdot (\vc{x} - \vc{x}^\prime)} , \label{eq:s}
\end{equation}
$\tau = t - t^\prime$, $\delta \tilde{\phi} = \delta \phi (\vc{k} , x^\prime, t)$ and $\delta \tilde{\vc{A}} = \delta \vc{A} (\vc{k}, x^\prime, t)$. The single particle characteristics ($\vc{x} - \vc{x}^\prime$) are quite complicated in the sheared electric and magnetic fields of a general neutral sheet configuration. In the VMID model, $S$ is computed numerically from Eq.~(\ref{eq:s}) for the exact single particle characteristics using the orbit integration technique described in Refs.~\onlinecite{daug:99} and \onlinecite{daug:03}. \

Equations (\ref{eq:dfs1}) and (\ref{eq:s}) are used to calculate the perturbed charge density ($\delta \rho = \sum_s q_s \delta n_s$ where $\delta n_s =  \int d^3v\, \delta f_s$) and current density ($\delta \vc{J} = \sum_s q_s \delta \vc{\Gamma}_s$ where $\delta \vc{\Gamma}_s = \int d^3v\, \vc{v} \delta f_s$), which are then used in Maxwell's equations in the Lorenz gauge
\begin{subequations}
\begin{eqnarray}
\nabla^2 \delta \phi - \frac{1}{c^2} \frac{\partial^2 \delta \phi}{\partial t^2} &=& - 4\pi \delta \rho , \label{eq:max1} \\ 
\nabla^2 \delta \vc{A} - \frac{1}{c^2} \frac{\partial^2 \delta \vc{A}}{\partial t^2} &=& - \frac{4\pi}{c} \delta \vc{J} . \label{eq:max2}
\end{eqnarray}
\end{subequations}
The resulting set of three coupled integro-differential equations for $\delta \phi$ and $\delta \vc{A}$ are then solved numerically as an eigenvalue problem for the eigenvectors and dispersion relation using a finite element approach.\cite{daug:03} The boundary conditions on $\delta \phi$ and $\delta \vc{A}$ are that they asymptote to a constant value (zero) as $x = \pm \infty$. 

\subsection{Small gyroradius limit \label{sec:sgr}}

As a next level of approximation, gyroradius effects are neglected ($\rho_s \rightarrow 0$). In this case, the single particle characteristics are $\vc{x} - \vc{x}^\prime = v_\parallel \tau \hat{b} + \mathcal{O}(\rho_s)$, where $\hat{b} = \vc{B}_o/|\vc{B}_o|$, and Eq.~(\ref{eq:s}) reduces to 
\begin{equation}
S = \frac{i (\delta \phi - v_\parallel \delta A_\parallel/c)}{\omega - k_\parallel v_\parallel} + \mathcal{O}(\rho_s) \label{eq:srho}
\end{equation}
assuming $\Im \lbrace \omega \rbrace >0$. To lowest-order in this gyroradius expansion, Eq.~(\ref{eq:dfs1}) can then be written
\begin{equation}
\delta f_s = - \frac{q_s f_{so}}{T_s} \biggl[ \delta \phi - \frac{\vc{U}_s \cdot \delta \vc{A}}{c} - \biggl( \delta \phi - \frac{v_\parallel \delta A_\parallel}{c} \biggr) \frac{\omega - \vc{k} \cdot \vc{U}_s}{\omega - k_\parallel v_\parallel} \biggr]  .  \label{eq:dfs2} 
\end{equation}
Since the phase speed of the waves of interest are much smaller than $c$, we drop the displacement current terms in Maxwell's equations, so $\nabla^2 \delta \phi = - 4\pi \delta \rho$ and $\nabla^2 \delta \vc{A} = -(4\pi/c)\delta \vc{J}$.

Evaluating the velocity-space integrals for the perturbed density and putting the result into Gauss's law gives 
\begin{align} 
[\partial_x^2 &- (k^2 + \lambda_{D}^{-2})] \delta \phi = \sum_s \lambda_{Ds}^{-2} \biggl\lbrace - \frac{\vc{U}_s \cdot \delta \vc{A}}{c} \label{eq:dphi3} \\ \nonumber
&+ \frac{\omega - \vc{k} \cdot \vc{U}_s}{k_\parallel v_{Ts}} \biggl[ \delta \phi Z(w_s) - \frac{v_{Ts}}{c} \delta A_\parallel (1 + \zeta_s Z(w_s)) \biggr] \biggr\rbrace  ,
\end{align}
in which $\lambda_{Ds}^2 = \sum_s (4\pi q_s^2 n_{so})/T_s$ is the square of the Debye length for species $s$, $\lambda_{D}^{-2} = \sum_s \lambda_{Ds}^{-2}$ gives the total Debye length,
\begin{equation}
w_{s} \equiv \frac{\omega - k_\parallel U_{s,\parallel}}{k_\parallel v_{Ts}} \ \ \ \textrm{and} \ \ \ \zeta_s \equiv \frac{\omega}{k_\parallel v_{Ts}}  .
\end{equation}
The coordinate rotation $(x,y,z) \rightarrow (x,b,\eta)$ has been applied where $\hat{b} = b_y \hat{y} + b_z \hat{z}$ is parallel to the equilibrium magnetic field and $\hat{\eta} = \hat{x} \times \hat{b} = -b_z \hat{y} + b_y \hat{z}$. The plasma dispersion function\cite{frie:61} 
\begin{equation}
Z_n (w) = \frac{1}{\sqrt{\pi}} \int_{-\infty}^\infty dt \frac{t^n e^{-t^2}}{t - w} 
\end{equation}
has also been used where $Z(w) \equiv Z_0(w)$. Evaluating the integrals for the perturbed current and putting the result into  Amp\`{e}re's law gives 
\begin{align}
\nabla^2 & \delta \vc{A} = \sum_s \frac{2 \omega_{ps}^2}{c k_\parallel v_{Ts}^2} \biggl\lbrace i \delta B_x \frac{U_{s,\eta}}{c} \vc{U}_s \label{eq:vca} \\ \nonumber
&+ i \delta E_\parallel \biggl[ \vc{U}_s + \frac{(\omega - \vc{k} \cdot \vc{U}_s)}{k_\parallel} \biggl( \frac{\vc{U}_s}{v_{Ts}} Z(w_s) + Z_1 (w_s) \hat{b} \biggr) \biggr] \biggr\rbrace 
\end{align} 
where $\omega_{ps} = \sqrt{4\pi q_s^2 n_{so}/m_s}$ is the plasma frequency of species $s$. Note also that $\delta B_x = i (k_\parallel \delta A_\eta - k_\eta \delta A_\parallel)$ and $\delta E_\parallel = -i k_\parallel \delta \phi + i (\omega/c) \delta A_\parallel$. 

Explicit evolution equations for $\delta A_y$ and $\delta A_z$ can be obtained by projecting Eq.~(\ref{eq:vca}) along $\hat{y}$ and $\hat{z}$. However, more compact expressions can be obtained by projecting Eq.~(\ref{eq:vca}) in the direction along $\vc{B}_o$ \textit{at a resonant surface} 
\begin{equation}
\hat{b}^s = \vc{B}_o (x_s) / B_o (x_s) = \cos \theta \hat{y} - \sin \theta \hat{z}  
\end{equation}
as well as the direction that is perpendicular to both this and $\hat{x}$, which is simply $\hat{k}$: $\hat{b}^s \times \hat{k} = \hat{x}$. Thus, $(\hat{x}, \hat{b}^s, \hat{k})$ provides a convenient orthogonal coordinate system. Applying this to Eq.~(\ref{eq:vca}) gives 
\begin{align}
[\partial_x^2 & - (k^2 + V_{b^s})] \delta \hat{A} = i \delta E_\parallel \sum_s \frac{2 \omega_{ps}^2}{c k_\parallel v_{Ts}} \biggl[ \frac{\hat{\vc{b}}^s \cdot  \vc{U}_s}{v_{Ts}}  \label{eq:dab3}  \\ \nonumber
&+ \frac{(\omega - \vc{k} \cdot \vc{U}_s)}{k_\parallel v_{Ts}} \biggl( \frac{\hat{\vc{b}}^s \cdot \vc{U}_s}{v_{Ts}} Z (w_s) + \frac{k_\eta}{k} Z_1 (w_s) \biggr) \biggr]
\end{align}
and 
\begin{align}
(\partial_x^2 & - k^2 ) \delta A_k = \delta \hat{A} V_k + i \delta E_\parallel \sum_s \frac{2 \omega_{ps}^2}{c k_\parallel v_{Ts}} \biggl[ \frac{\hat{\vc{k}} \cdot  \vc{U}_s}{v_{Ts}}  \label{eq:dak3}  \\ \nonumber
&+ \frac{(\omega - \vc{k} \cdot \vc{U}_s)}{k_\parallel v_{Ts}} \biggl( \frac{\hat{\vc{k}} \cdot \vc{U}_s}{v_{Ts}} Z (w_s) + \frac{k_\parallel}{k} Z_1 (w_s) \biggr) \biggr]
\end{align}
in which 
\begin{equation}
\vc{V} = \sum_s \frac{2 \omega_{ps}^2}{c^2} \frac{k}{k_\parallel} \frac{U_{s,\eta} \vc{U}_s}{v_{Ts}^2}  , \label{eq:vdef}
\end{equation}
$V_{b^s} = \hat{\vc{b}}^s \cdot \vc{V}$ and $V_{k} = \hat{\vc{k}} \cdot \vc{V}$.
Here, $\delta A_k = \vc{k} \cdot \delta \vc{A}$,
\begin{equation}
\delta \hat{\vc{A}} = \vc{b}^s \cdot \delta \vc{A} =  (k_\eta \delta A_\parallel - k_\parallel \delta A_\eta)/k = i \delta B_x/k ,
\end{equation}
and
\begin{equation}
i \delta E_\parallel =  k_\parallel \delta \phi - \frac{\omega}{c} \biggl( \frac{k_\eta}{k} \delta \hat{A} + \frac{k_\parallel}{k} \delta A_k \biggr)   .
\end{equation}

The set of three coupled equations (\ref{eq:dphi3}), (\ref{eq:dab3}) and (\ref{eq:dak3}) provide a closed description for the small gyroradius limit. 

\subsection{Decoupled magnetic field approximation\label{sec:decoupled}}

Although $\delta E_\parallel$ in Eq.~(\ref{eq:dab3}) depends on $\delta \phi$, $\delta \hat{A}$ and $\delta A_k$, consideration of two-scale features of a tearing layer can be used to justify an approximate decoupling of $\delta \hat{A}$ from $\delta \phi$ and $\delta A_k$. In the outer region, far from a resonant surface, magnetic flux is frozen into the plasma so the parallel electric field vanishes $\delta E_\parallel \rightarrow 0$. In this ideal MHD like outer region, the right side of Eq.~(\ref{eq:dab3}) is negligible and $\delta \hat{A}$ is the only remaining dependent variable. In the inner region near a resonant surface, $k_\parallel \ll k$, which implies
\begin{equation}
i \delta E_\parallel = k_\parallel \delta \phi - \frac{\omega}{c} \frac{k_\eta}{k} \delta \hat{A} - \frac{\omega}{c} \frac{k_\parallel}{k} \delta A_k \simeq - \frac{\omega}{c} \delta \hat{A} , \label{eq:epaprx}
\end{equation} 
and
\begin{equation}
\hat{\vc{b}}^s \cdot \vc{U}_s = b_\parallel^s U_{s,\parallel} + b_\eta^s U_{s,\eta} = \frac{k_\eta}{k} U_{s,\parallel} - \frac{k_\parallel}{k} U_{s,\eta} \simeq U_{s,\parallel}
\end{equation}
so 
\begin{equation}
\frac{k}{k_\eta} \frac{\hat{\vc{b}}^s \cdot \vc{U}_s}{v_{Ts}} Z(w_s) + Z_1(w_s) \simeq 1 + \zeta_s Z(w_s) .
\end{equation}
If we also note that the first term on the right side of Eq.~(\ref{eq:dab3}) is small compared to the other two, we are left with 
\begin{align}
[\partial_x^2 &- (k^2 + V_{b^s})] \delta \hat{A} =   \label{eq:dahat} \\ \nonumber 
&- \delta \hat{A} \sum_s \frac{2 \omega_{ps}^2}{c^2} \frac{(\omega - \vc{k} \cdot \vc{U}_s)}{k_\parallel v_{Ts}} \zeta_s [1 + \zeta_s Z(w_s)]  .
\end{align}
to describe the inner region. 

Although Eq.~(\ref{eq:dahat}) was derived for the inner region, it is expected to hold in the outer region as well. This is because the left side is the ideal MHD outer region equation, and in the outer region the right side is negligibly small since it is proportional to $\omega^2/(kv_{Ts})^2 \ll 1$. The right side contributes only in the inner region where $k_\parallel$ is small. Thus, Eq.~(\ref{eq:dahat}) is an approximate equation spanning the layer that is decoupled from $\delta A_k$ and $\delta \phi$. 
The validity of this scale separation will be evaluated in Sec.~\ref{sec:const_psi}. 

\subsection{Harris Equilibrium \label{sec:equil}}

We utilize the following properties of the Harris equilibrium\cite{harr:62}
\begin{subequations}
\begin{eqnarray}
\vc{B}_o &=& B_{oy} \hat{y} + \bar{B}_{oz} \tanh (x/\lambda) \hat{z} , \\
\bar{n} &=& \bar{n}_o \sech^2 (x/\lambda) , \label{eq:nsharris} \\
\bar{B}_{oz}^2 &=& 8\pi \bar{n}_o (T_e + T_i) , \label{eq:bozharris} \\
\vc{U}_s &=& - 2cT_s/(q_s \bar{B}_{oz} \lambda) \hat{y}  \label{eq:harris_us}
\end{eqnarray}
\end{subequations}
where an additional uniform background density ($n_b$) for each species is included, $n(x)=\bar{n}(x) + n_b$. Noting that $U_{s,\eta} = -b_z U_s$, Eq.~(\ref{eq:vdef}) reduces to
\begin{equation}
\vc{V} = - \frac{4\pi \bar{n} e^2}{c^2} \biggl(\frac{U_i}{T_i} \biggr)^2 (T_e + T_i) \frac{k B_z}{k_\parallel B_o} \hat{y} , \label{eq:vharris1}
\end{equation}
and Amp\'{e}re's law gives
\begin{equation}
B_z^{\prime \prime} = - \frac{4\pi e}{c} \frac{U_i}{T_i} (T_e + T_i) \bar{n}^\prime , \label{eq:bzpp1}
\end{equation}
where ``prime'' denotes a derivative. 
Taking a derivative of Eq.~(\ref{eq:nsharris}) and utilizing Eqs.~(\ref{eq:bozharris}) and (\ref{eq:bzpp1}) shows
\begin{equation}
B_{z}^{\prime \prime} = - \frac{4\pi \bar{n} e^2}{c^2}  \biggl( \frac{U_i}{T_i} \biggr)^2 (T_e + T_i) B_z  .
\end{equation}
Putting this into Eq.~(\ref{eq:vharris1}) provides
\begin{equation}
\vc{V} = \frac{F^{\prime \prime}}{F} \frac{k}{k_z} \hat{y}   \label{eq:vharris}
\end{equation}
where $F = \vc{k} \cdot \vc{B}_o$. The individual components of interest are then
\begin{equation}
V_{b^s} = \frac{F^{\prime \prime}}{F} = - \frac{2}{\lambda^2} \frac{\sech^2 (x/\lambda ) \tanh (x/\lambda )}{\tanh (x/\lambda) - \tanh (x_s/\lambda)} ,  \label{eq:vbs}
\end{equation}
and $V_k = \tan \theta V_{b^s}$.

\section{Numerical Solutions \label{sec:numsol}} 

In this section, the dispersion relation computed from the VMID model of Sec.~\ref{sec:vlasov} is compared with that computed from the approximate layer equation of Sec.~\ref{sec:decoupled} [Eq.~(\ref{eq:dahat})]. 
The VMID computations were calculated using the method and code described in Refs.~\onlinecite{daug:99} and \onlinecite{daug:03}. 
Solutions of Eq.~(\ref{eq:dahat}) were obtained by numerically integrating from $x=\pm \infty \rightarrow x_s$, and determining the dispersion relation from the requirement that $\delta \hat{A}^\prime(x_s^+) = -\delta \hat{A}^\prime(x_s^-)$.

\subsection{Growth rate profiles} 

\begin{figure}
\includegraphics[width=8.5cm]{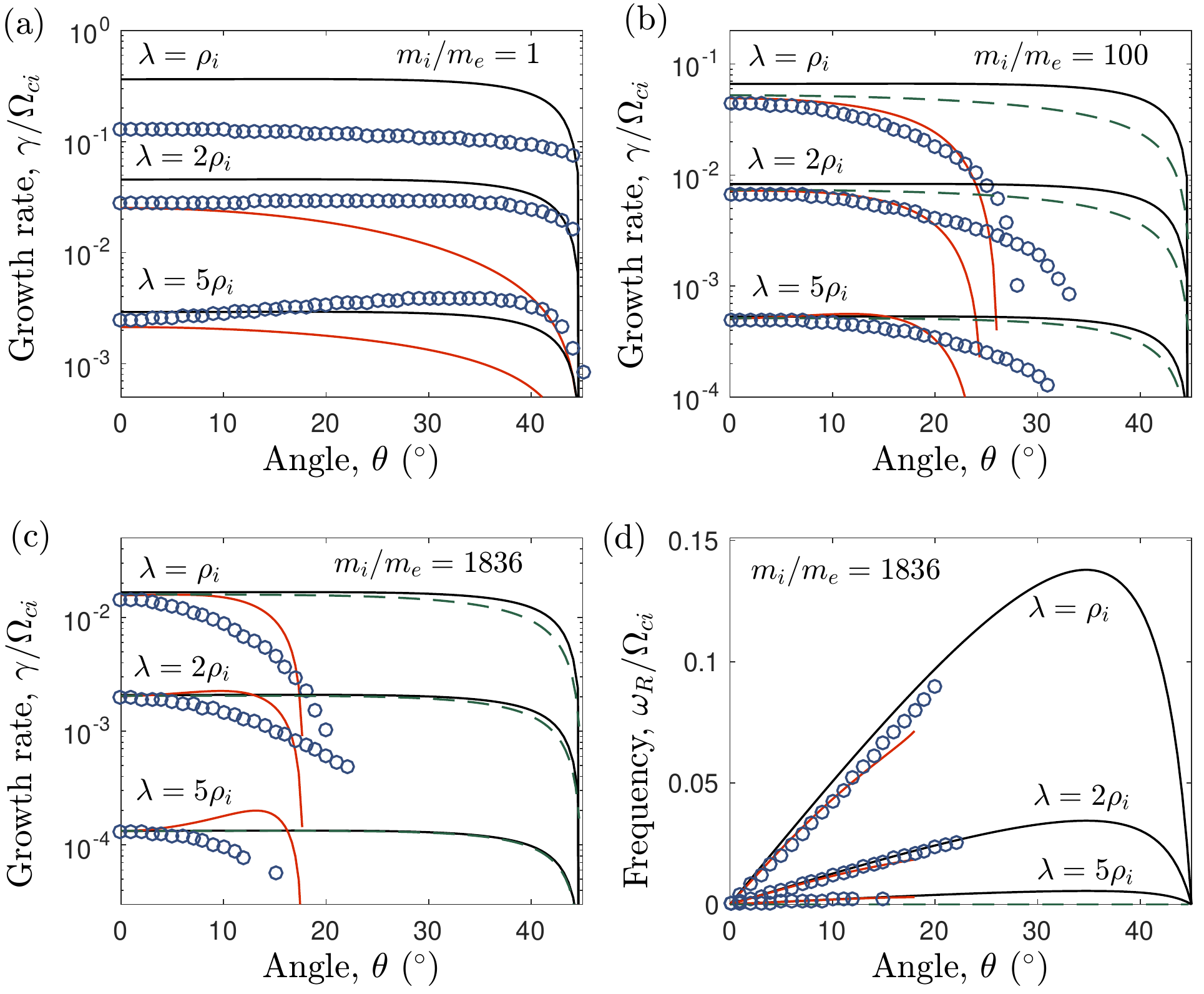}
\caption{(a)-(c) Growth rates and (d) real frequency computed from the VMID model (circles), the decoupled layer equation from Eq.~(\ref{eq:dahat}) (red line),  Eq.~(\ref{eq:dahat}) with $\vc{U}_s = 0$ (dashed line) and the Drake and Lee theory from Eq.~(\ref{eq:gdrake}) (black line).  Other parameters include: $T_i/T_e = 1$,  $k\lambda = 0.4$, $n_b = 0.3n_o$, $B_{oy}/\bar{B}_{oz} = 1$.}
\label{fg:gam_gal}
\end{figure}

The VMID solutions shown in Fig.~\ref{fg:gam_gal}c demonstrate that at realistic mass ratio ($m_i/m_e = 1836$), the growth rate is a maximum at $\theta = 0^\circ$, corresponding to the usual parallel mode at the center of the current sheet, and monotonically decreases with increasing angle of obliquity. The growth rate falls off slowly with angle until near $\theta = 20^\circ$, where it rapidly decreases, and quickly reaches stability for larger angles. These general features of the growth rate profiles hold at each of the current sheath widths considered ($\lambda = \rho_i$, $2\rho_i$ and $5\rho_i$), but details of the slope of the fall off differ in each case. The real frequency of the instability is to a good approximation linear over the entire range of unstable angles at each sheet width. 

Figure~\ref{fg:gam_gal}b shows results of analogous simulations, but at lower mass ratio ($m_i/m_e = 100$). Similar trends in the growth rate as for the proton-electron mass ratio are observed, including the maximum growth rate at $\theta=0^\circ$ and a monotonic decrease of the growth rate. The primary differences in this comparison are that the unstable range of wavenumber extends further out on the current sheet (to $\theta \simeq 30^\circ$), and that the values of the growth rate in units of $\Omega_{ci}$ take larger values at lower mass ratio. Here, $\Omega_{ci} = e\bar{B}_{oz}/(m_ic)$ is the ion gyrofrequency based on the asymptotic magnetic field strength $\bar{B}_{oz}$. In the latter comparison, an additional consideration is the mass dependence of the growth rate units being displayed. Accounting for this, $1836/100=18.36$, demonstrates that in dimensional units ($s^{-1}$) the growth rate is higher at larger mass ratio. 

Figure~\ref{fg:gam_gal}a shows that at mass ratio of unity $m_i/m_e=1$, qualitative differences in the growth rate profiles are observed. The most apparent is that the growth rate is positive and near constant over the entire range of angles. This suggests that at unity mass ratio linear tearing modes would be excited throughout the entire current sheet; a result that has also been observed in kinetic simulations.~\cite{yin:08} Another apparent difference is that for the larger current sheet widths ($\lambda = 5 \rho_i$ and, to a lesser extent, $\lambda = 2 \rho_i$), the most unstable mode is oblique. 

Each of the panels in Fig.~\ref{fg:gam_gal} also show the predictions of the approximate layer equation from Eq.~(\ref{eq:dahat}). Comparison with the Vlasov simulation curves shows that, despite the severity of some the approximations made in Secs.~\ref{sec:sgr} and \ref{sec:decoupled}, these solutions capture key features of the growth rate profiles and also give a fair quantitative prediction. The key agreement is that at large mass ratios ($m_i/m_e =100$ and 1836) the range of unstable angles is well approximated by this solution. From the perspective of understanding how linear tearing modes will contribute to the overall reconnection dynamics, this is a key feature because it determines the locations in the current layer that are expected to be unstable.  
It is also observed that at large mass ratios, Eq.~(\ref{eq:dahat}) accurately predicts the growth rate of the central mode at $\theta = 0^\circ$, and generally predicts a flat profile. One distinction observed at $\lambda=5\rho_i$ and $m_i/m_e = 1836$ is that Eq.~(\ref{eq:dahat}) predicts the most unstable mode is oblique (near the cutoff), whereas the Vlasov solutions predict a monotonically decreasing profile. Finally, much poorer agreement between Eq.~(\ref{eq:dahat}) and the Vlasov solutions is observed at unity mass ratio. With the exception of the larger current sheet widths and angles near the center of the current sheet, the agreement is poor. 
This is expected because, for a pair plasma, the electron gyroradius becomes comparable to the gradient scale of the current sheet. 

The observed stabilization at sufficiently oblique angles is associated with the diamagnetic drift $\vc{U}_s$. This is confirmed by the dashed lines in Fig.~\ref{fg:gam_gal}, which show solutions of Eq.~(\ref{eq:dahat}) with $\vc{U}_s=0$. These show that in the absence of the diamagnetic drift, the theory predicts positive and nearly constant growth rates across the entire current sheet. Recall from Eq.~(\ref{eq:harris_us}) that the diamagnetic drift in the Harris equilibrium is associated with the density gradient, since the temperature is uniform. 
Cowley, Kulsrud and Hahm previously studied the influence of temperature gradients.\cite{cowl:86}
It is also interesting to note that for force-free equilibria, which do not have diamagnetic drifts, tearing modes have been observed to span the current sheet~\cite{liu:13,akca:16}
The results of Eq.~(\ref{eq:dahat}) in the absence of diamagnetic drift agree well with the predictions of the standard kinetic tearing mode theory of Drake and Lee \cite{drak:77}. 
In Sec.~\ref{sec:const_psi} it is shown that the reason for the absence of the drift stabilization of oblique modes in the theory is a broadening of the inner layer at oblique angles, which causes the boundary layer approximation that the theory utilizes to break down. 
Because the full numerical solution of Eq.~(\ref{eq:dahat}) accurately captures the essential feature of the unstable region of the current layer, the remainder of this work will focus on this equation.

\subsection{Vector potential and phase shifts}

\begin{figure}
\includegraphics[width=8.5cm]{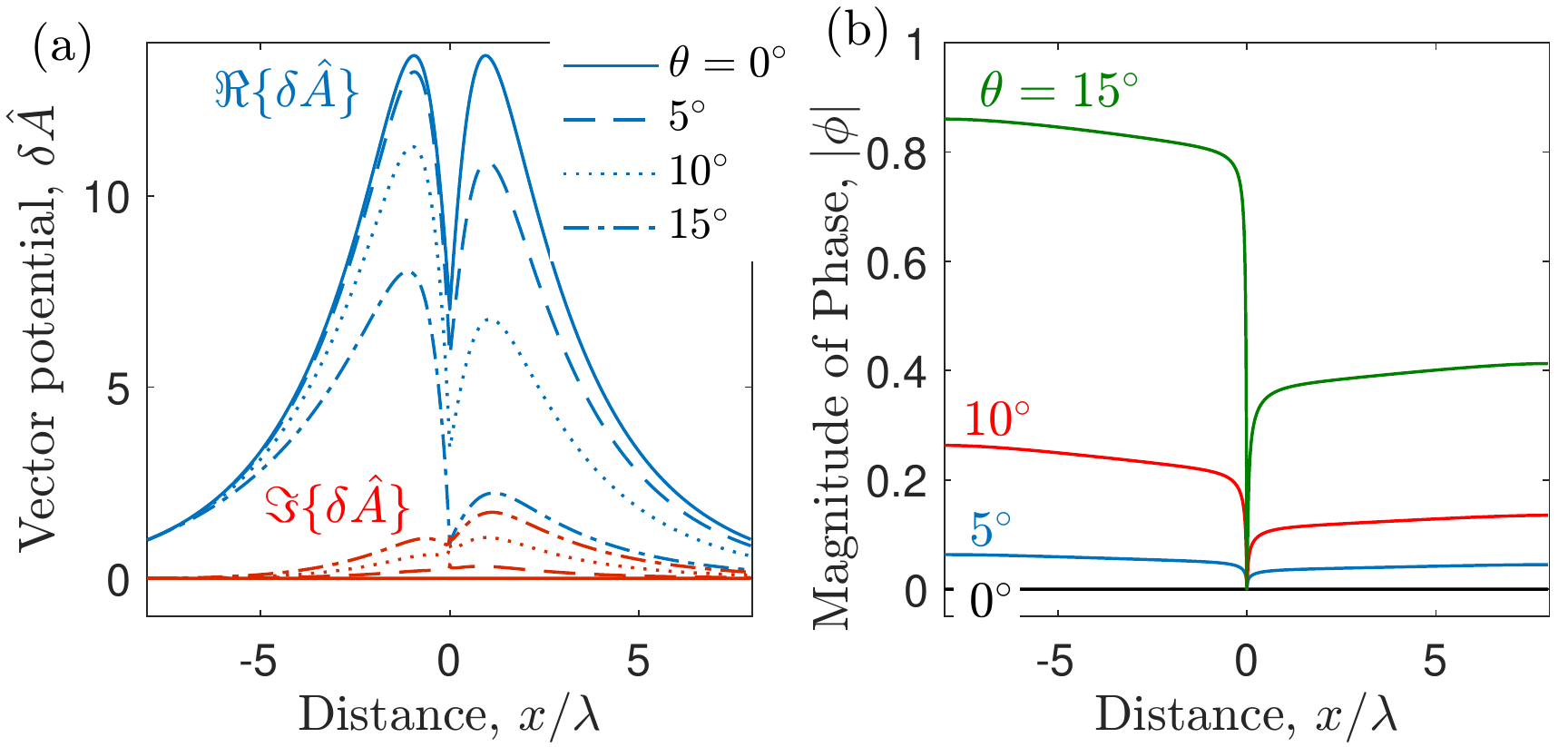}
\caption{(a) Real (blue) and imaginary (red) components of the vector potential computed from solutions of Eq.~(\ref{eq:dahat}) for plasma parameters $\hat{n}_b = 0.3$, $\lambda = 5 \rho_i$, $k\lambda = 0.4$, $\bar{B}_{oz}/B_{oy} = 1$, $T_i = T_e$.  (b) The magnitude of the associated phase angle.  }
\label{fg:phase}
\end{figure}

Figure~\ref{fg:phase}a shows profiles of the real and imaginary components of the vector potential $\delta \hat{A}(x)$ across the current sheet for modes centered at resonant surfaces corresponding to $\theta =0^\circ$, $5^\circ$, $10^\circ$ and $15^\circ$. 
This solution assumed a proton-electron mass ratio $m_i/m_e=1836$ and the other parameters chosen were the same as in Fig.~\ref{fg:gam_gal}c. The eigenfunction corresponding to the mode at the central resonant surface ($\theta = 0^\circ$) is purely real and is symmetric about $x=0$. This is the common MHD solution of the outer region familiar from the seminal Furth, Killeen and Rosenbluth (FKR)\cite{furt:63} analysis. Oblique modes are centered at resonant surfaces off-center of the current sheet, and the most pronounced feature of the eigenfunctions is the formation of an asymmetry about the resonant surface. This asymmetry was also observed in our previous reduced MHD analysis of oblique tearing modes (see figure 7 of Ref.~\onlinecite{baal:12}), and is associated with ideal MHD outer region for off-center modes. 

A distinguishing feature of the kinetic solutions, compared to MHD solutions, is that the eigenfunctions of the oblique modes are complex. Figure~\ref{fg:phase} shows that all off-center modes ($\theta > 0$) have a finite $\Im \lbrace \delta \hat{A}(x) \rbrace$, and that their profile is also asymmetric with respect to the resonant surface. 
The values of this component are negative (the plot shows $\Im \lbrace - \delta \hat{A} \rbrace$). 
Figure~\ref{fg:phase}b shows that if the complex vector potential is represented as $\delta \hat{A}(x) = |\delta \hat{A}(x)| \exp[i\phi(x)]$, there is a phase shift of the eigenfunction across the current sheet for oblique modes. The phase factor asymptotes to a constant value far from the center of the current sheet, and vanishes at the resonant surfaces. This phase shift is observed to increase with increasing angle of obliquity. This phase factor is also a distinctly kinetic feature of the eigenfunctions. 

Boundary layer theory assumes that the outer region is described by ideal MHD, and does not account for a complex vector potential. 
Thus, one may expect the larger imaginary component of $\delta \hat{A}$ at oblique angles to be associated with the observed breakdown of the analytic theories. 
However, we find that the dominant effect is a breakdown of the boundary layer scale separation, rather than the complex eigenfunction. 
This aspect will be discussed further in Sec.~\ref{sec:const_psi}.

\section{Relation to previous theories\label{sec:compare}} 

\subsection{Drake and Lee \label{sec:dl}}

The seminal kinetic tearing mode theory of Drake and Lee (DL)\cite{drak:77} applies a boundary-layer analysis that considers only electrons in the inner region. A dispersion relation is obtained by matching the inner layer solution with an outer layer solution through a matching parameter called the tearing stability index 
\begin{equation}
\Delta^\prime \equiv [\delta \hat{A}^\prime (x_s^+) - \delta \hat{A}^\prime (x_s^-)]/\delta \hat{A} (x_s)
\end{equation}
which is provided by an independent ideal MHD solution. Generalizing their inner region equation to include ions gives
\begin{equation}
\delta \hat{A}^{\prime \prime} = - \delta \hat{A} \sum_s \frac{2\omega_{ps}^2}{c^2} \frac{(\omega - \vc{k} \cdot \vc{U}_s)}{k_\parallel v_{Ts}} \zeta_s Z_1(\zeta_s) . \label{eq:dlinner}
\end{equation}
Equation~(\ref{eq:dlinner}) agrees with the inner region of our decoupled layer equation (\ref{eq:dahat}) in the limit that $w_s \rightarrow \zeta_s$, which implies $\omega/k_\parallel \gg U_{s,\parallel}$. Note, $Z_1(\zeta_s) = Z^\prime(\zeta_s)/2=1+\zeta_sZ(\zeta_s)$.

\begin{figure}
\includegraphics[width=7.5cm]{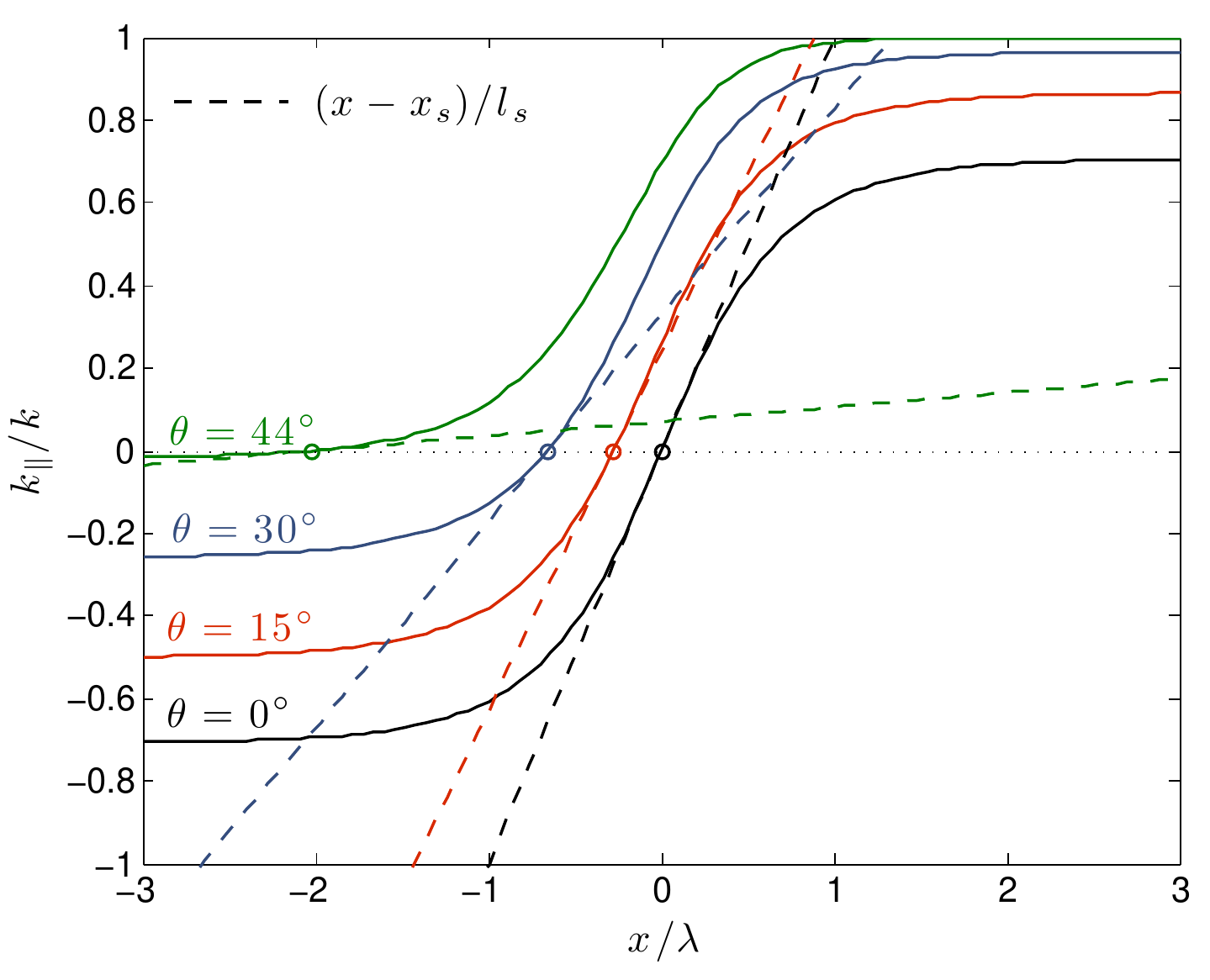}
\caption{Parallel wavenumber of modes associated with different resonant surfaces ($\theta = 0^\circ$, $15^\circ$, $30^\circ$ and $44^\circ$) across the current sheet using Eq.~(\ref{eq:kpar_expan}) (solid lines) and the linear expansion using Eq.~(\ref{eq:ls}) (dashed lines). These curves were obtained assuming $B_{oy}/\bar{B}_{oz} = 1$. }
\label{fg:kpar}
\end{figure}

The DL dispersion relation follows from applying the constant-$\psi$ approximation. In this approximation, $\delta \hat{A}$ on the right side of Eq.~(\ref{eq:dlinner}) is assumed constant across the inner layer. Dividing by this and integrating across the layer gives $\int dx\, \delta \hat{A}^{\prime \prime}/\delta \hat{A} (x_s) \simeq \Delta^\prime$, and 
\begin{equation}
\Delta^\prime \simeq  \sum_s \frac{(\omega - \vc{k} \cdot \vc{U}_s)}{\omega} \int_{-\infty}^\infty dx \frac{\omega_{ps}^2}{c^2}  \zeta_s^2 Z^\prime (\zeta_s) .  \label{eq:drake1}
\end{equation}
The full solution of $k_\parallel$ for the Harris equilibrium described in Sec.~\ref{sec:equil} is 
\begin{equation}
\label{eq:kpar_expan}
\frac{k_\parallel}{k} = \cos \theta \frac{\mu + \tanh (x/\lambda)}{\sqrt{ B_{oy}^2/\bar{B}_{oz}^2 + \tanh^2 (x/\lambda)}} .
\end{equation}
To integrate the right side of Eq.~(\ref{eq:drake1}) an additional assumption is applied, where the space-dependent functions $k_\parallel$ and $\omega_{ps}$ are expanded about the resonant surface location: $k_\parallel \simeq k (x-x_s) /l_s$, and $\omega_{ps} = \omega_{ps}(x_s)$, where
\begin{equation}
l_s =  \frac{k}{k_\parallel^\prime (x_s)} = \frac{\lambda B_{oy}/ \bar{B}_{oz}}{\cos^2\theta (1 - \mu^2)}  . \label{eq:ls}
\end{equation}
Figure~\ref{fg:kpar} shows a plot of the full space-dependent solution of Eq.~(\ref{eq:kpar_expan}) along with the linear expansion at several resonant surface locations. This shows that the width over which the linear exapansion is accurate depends on the resonant surface location.

In Ref.~\onlinecite{drak:77}, the integral in Eq.~(\ref{eq:kpar_expan}) is evaluated by applying the above approximations.
The result is 
\begin{equation}
\label{eq:dl_int}
\int_{-\infty}^{\infty} dx\, \zeta_s^2 Z^\prime (\zeta_s) = - \frac{2i \sqrt{\pi} \omega l_s}{kv_{Ts}} .
\end{equation}
With this, Eq.~(\ref{eq:drake1}) provides the dispersion relation
\begin{equation}
\Delta^\prime = -2i\sqrt{\pi} \sum_s \frac{l_s (\omega -\vc{k} \cdot \vc{U}_s)}{k v_{Ts} d_s^2} \label{eq:dl_disp}
\end{equation}
where $d_s\equiv c/\omega_{ps}$ is the skin depth associated with species $s$. 
Rearranging, the real frequency and growth rate are 
\begin{equation}
\label{eq:dl_disp}
\omega_{\textrm{DL}} = \frac{\sum_s \vc{k} \cdot \vc{U}_s/(v_{Ts} d_s^2)}{\sum_s 1/ (v_{Ts} d_s^2)} + i \frac{k \Delta^\prime}{2 \sqrt{\pi} l_ s \sum_s 1/(v_{Ts} d_s^2)}  .
\end{equation}

Applying the expressions for a Harris sheet with a background density from Sec.~\ref{sec:equil} provides $\omega = \omega_r + i \gamma$, where
\begin{equation}
\omega_r = \frac{k_y U_e (1 - \alpha T_i/T_e)}{(1 + \alpha) [1 + \hat{n}_b/(1-\mu^2)]}  \label{eq:wdrake}
\end{equation}
is the real frequency of the tearing mode, and 
\begin{equation}
\gamma = \frac{k \Delta^\prime v_{Te} d_{e}^2|_{x_s}}{2 \sqrt{\pi} l_s (1+\alpha)}  \label{eq:gdrake}
\end{equation}
is the growth rate. Here, $\hat{n}_b \equiv n_b/\bar{n}_o$ and $\alpha \equiv \sqrt{T_em_e/(T_im_i)}$. For the plots in Fig.~\ref{fg:gam_gal}, which show $\omega/\Omega_{ci}$, it is convenient to note that $k_y U_e = \Omega_{ci} k\lambda \sin \theta (T_e/T_i) (\rho_i/\lambda)^2$ and $v_{Te} d_{e}^2|_{x_s} = \Omega_{ci} \rho_i^3 \alpha (1+T_e/T_i)/(1-\mu^2 + \hat{n}_b)$. 

\begin{figure}
\includegraphics[width=8.5cm]{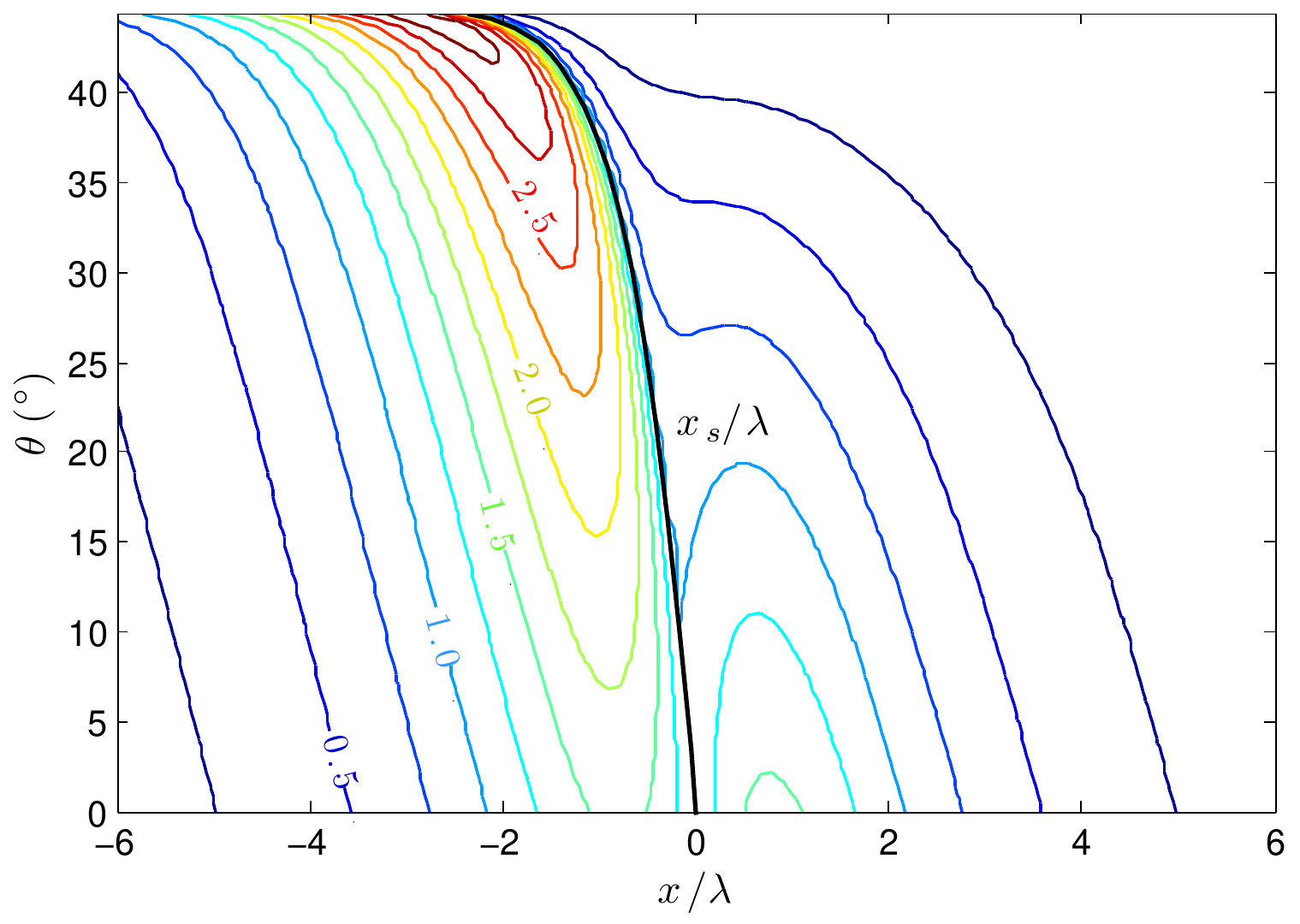}
\caption{Contours of constant $\delta \hat{A}$ from a numerical solution of the FKR outer region from Eq.~(\ref{eq:fkr_outer}). The black line shows the location of the resonant surface.  The parameters $k \lambda = 0.5$ and $B_{oy}/\bar{B}_{oz} = 1$ were chosen.}
\label{fg:dacont}
\end{figure}

The tearing stability index $\Delta^\prime$ must be supplied from a model for the outer region of the boundary layer. 
This is typically assumed to obey ideal MHD (FKR)~\cite{furt:63}
\begin{equation}
\delta \hat{A}^{\prime \prime} - (k^2 + V_{b^s}) \delta \hat{A} = 0 . \label{eq:fkr_outer}
\end{equation}
Recall that $V_{b^s} = F^{\prime \prime}/F$ is provided in Eq.~(\ref{eq:vbs}). 
Figure~\ref{fg:dacont} shows the solution of Eq.~(\ref{eq:fkr_outer}) throughout the domain of possible resonant surfaces.
One prominent feature is that $\delta \hat{A}(x)$ is an even function of $x$ only for $\theta = 0$. For off center modes, the characteristic peak is much higher on the low magnetic field side of the current sheet ($x<x_s$) than on the high magnetic field side ($x>x_s$). A model for $\Delta^\prime$ based on Eq.~(\ref{eq:fkr_outer}), which considers oblique modes, was developed in Ref.~\onlinecite{baal:12} 
\begin{equation}
\Delta^\prime_H \simeq \frac{2}{\lambda} \biggl( \frac{1 + \mu^2}{k\lambda} - k \lambda \biggr) . \label{eq:dph}
\end{equation}
This was validated against numerical solutions of Eq.~(\ref{eq:fkr_outer}) over a broad range of conditions (see Fig.~3 of Ref.~\onlinecite{baal:12}), and a similar validation is shown in Fig.~\ref{fg:dacomp}b. It was used to evaluate $\Delta^\prime$ for the Drake and Lee theory curves shown in Fig.~\ref{fg:gam_gal}. 

Figure~\ref{fg:gam_gal} shows that the solution of the DL theory, obtained this way, accurately predicts the growth rate of parallel ($\theta = 0^\circ$) modes; the only exception being thin current sheets at unity mass ratio. 
However, the figure also shows that theory does not accurately capture oblique modes, particularly the stabilization. It is interesting to notice that Eq.~(\ref{eq:dahat}) does capture the range of unstable modes, and that it is very closely related to the fundamental equation that the DL boundary layer theory is based on. The only difference is that the argument of the $Z$-function in Eq.~(\ref{eq:dahat}) is $w_s$, whereas it is $\zeta_s$ in DL theory. We have also solved numerically a version of Eq.~(\ref{eq:dahat}) with $Z(w_s)$ replaced by $Z(\zeta_s)$, and have found that the solutions are very close (they also capture the stabilization of oblique modes). 
Thus, the discrepancy with DL theory cannot be explained by this difference.  

\subsection{Galeev et al}

\begin{figure}
\includegraphics[width=8.5cm]{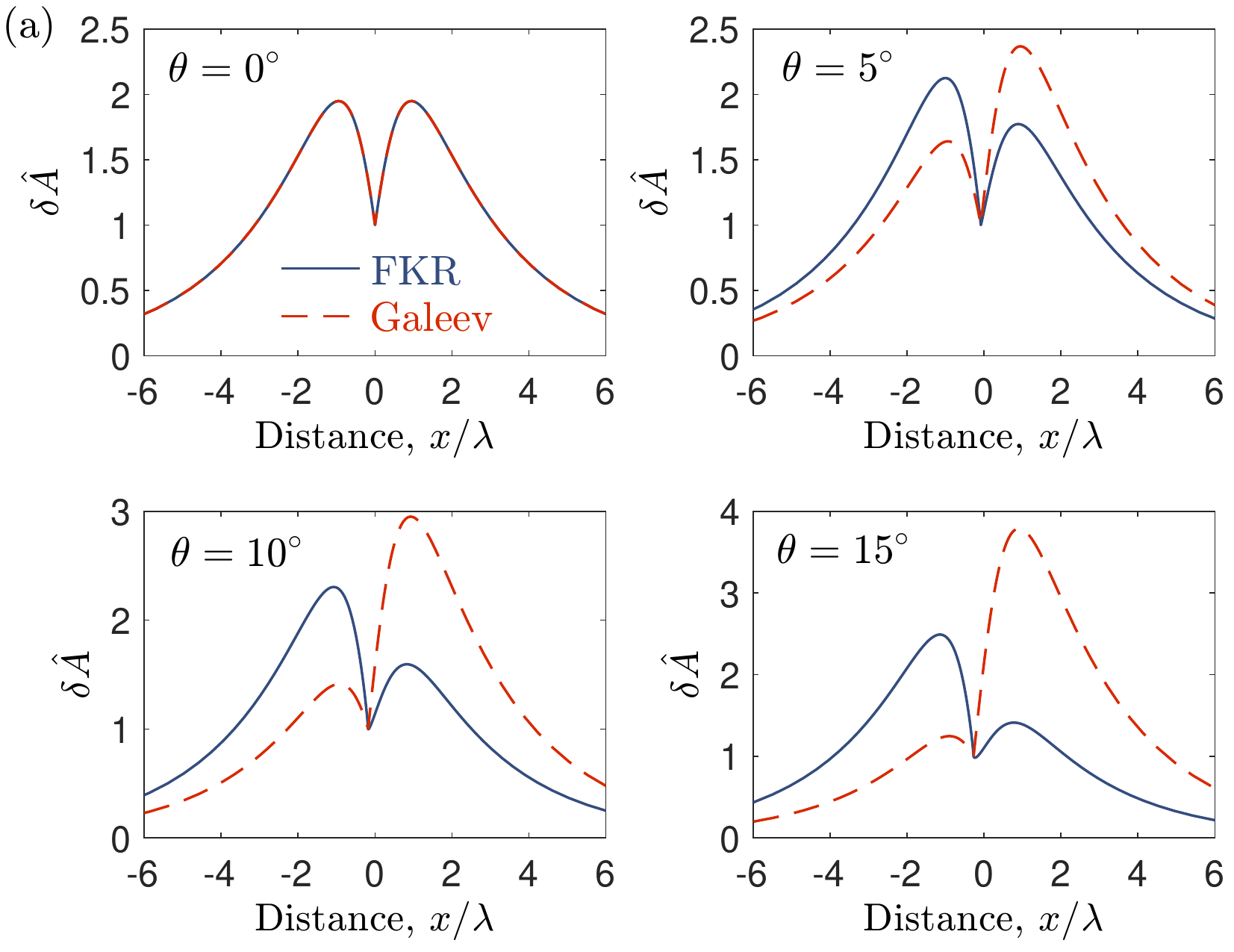}
\includegraphics[width=8.0cm]{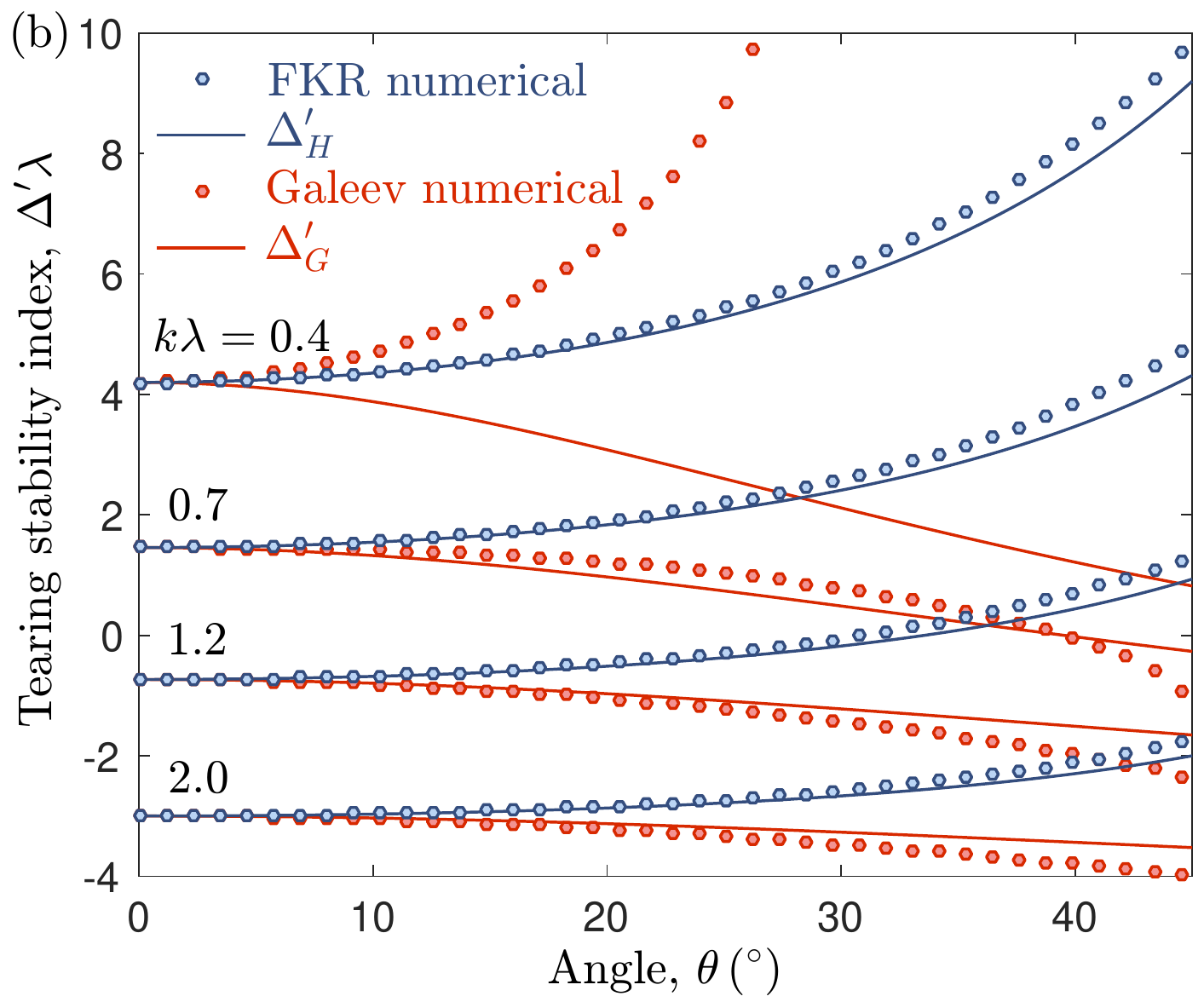}
\caption{(a) Numerical solutions of $\delta \hat{A}$ from the ideal MHD outer region from Eq.~(\ref{eq:fkr_outer}) (solid lines) and Galeev outer region from Eq.~(\ref{eq:gal_outer}) (dashed lines) at four values of the angle of obliquity. Here, $k \lambda = 0.4$, $B_{oy}/\bar{B}_{oz} = 1$ have been chosen. (b) $\Delta^\prime$ obtained from numerical solutions of the FKR outer region Eq.~(\ref{eq:fkr_outer}) (blue circles), the Galeev outer region Eq.~(\ref{eq:gal_outer}) and the approximate solutions from Eqs.~(\ref{eq:dph}) and (\ref{eq:dpg}) respectively (solid lines). }
\label{fg:dacomp}
\end{figure}

Previous researchers\cite{kuzn:85,gale:86,glad:90} have studied oblique collisionless tearing modes in the context of Earth's magnetopause, and have obtained a different result than Drake and Lee. These papers keep a first-order finite gyroradius correction, but the small gyroradius limit of their equations can be derived from Eq.~(\ref{eq:dfs2}) by applying the following assumptions: (1) $k_\parallel \ll k$ and (2) $U_s/v_{Ts} \ll1$. These imply 
\begin{equation}
\frac{\vc{U}_s \cdot \delta \vc{A}}{c} \ll \frac{(\omega - \vc{k} \cdot \vc{U}_s)}{k_\parallel c} \delta A_\parallel \ \ \ \textrm{and} \ \ \ w_s \simeq \zeta_s ,
\end{equation}
so Eq.~(\ref{eq:dphi3}) reduces to
\begin{align}
(\partial_x^2 &- k^2 - \lambda_D^{-2}) \delta \phi = \label{eq:phigaleev} \\ \nonumber
& \sum_s \lambda_{Ds}^{-2} \frac{(\omega - \vc{k} \cdot \vc{U}_s)}{k_\parallel v_{Ts}} \biggl[ \delta \phi Z(\zeta_s) - \frac{v_{Ts}}{c} Z_1 (\zeta_s) \delta A_\parallel \biggr] .
\end{align}
For Amp\'{e}re's equation, it is convenient to first rearrange Eq.~(\ref{eq:vca}) to the form
\begin{align}
\nabla^2 \delta \vc{A} &= \sum_s \frac{2 \omega_{ps}^2}{c v_{Ts}^2} \biggl\lbrace \biggl( \delta \phi - \frac{\vc{U}_s \cdot \delta \vc{A}}{c} \biggr) \vc{U}_s  \label{eq:vca2} \\ \nonumber
&+ \frac{(\omega - \vc{k} \cdot \vc{U}_s)}{k_\parallel} \biggl[ \delta \phi \biggl( \frac{\vc{U}_s}{v_{Ts}} Z(w_s) + Z_1 (w_s) \hat{b} \biggr) \\ \nonumber
&- \delta A_\parallel \frac{v_{Ts}}{c} \biggl( \frac{\vc{U}_s}{v_{Ts}} (1 + \zeta_s Z(w_s)) + \zeta_s Z_1 (w_s) \hat{b} \biggr) \biggr] \biggr\rbrace .
\end{align}
The $k_\parallel \ll k$ assumption implies $\delta \vc{A} \simeq \delta \hat{A} \hat{\vc{b}}^s$ and $\hat{\vc{b}}^s \simeq \hat{b} \simeq \hat{z}$. Applying these along with assumption (2), the $\hat{\vc{b}}^s$ projection of Eq.~(\ref{eq:vca2}) reduces to 
\begin{equation}
(\partial_x^2 - k^2 - V_{\textrm{ad}} ) \delta \hat{A} = i \delta E_\parallel \sum_s \lambda_{Ds}^{-2} \frac{(\omega - \vc{k} \cdot \vc{U}_s)}{k_\parallel^2 c} Z_1 (\zeta_s)   \label{eq:agaleev}
\end{equation}
in which 
\begin{equation}
V_\textrm{ad} = - \sum_s \frac{2 \omega_{ps}^2}{c^2} \frac{(\hat{\vc{b}}^s \cdot \vc{U}_s)^2}{v_{Ts}^2} 
\end{equation}
is what Galeev calls the adiabatic interaction term. 

Equations~(\ref{eq:phigaleev}) and (\ref{eq:agaleev}) are the two basic equations that this line of previous collisionless oblique tearing mode research\cite{kuzn:85,gale:86,glad:90} worked from.\cite{note1:gale} This set of two coupled equations is substantially different than any of the approximate equations derived in Sec.~\ref{sec:be}. The most significant difference is in the description of the outer region. For an ideal MHD outer region, the $V_\textrm{ad}$ is replaced by $V_{b^s}$, which reduces to Eq.~(\ref{eq:vbs}) for a Harris equilibrium. However, the adiabatic response used in Eq.~(\ref{eq:agaleev}) is 
\begin{equation}
V_\textrm{ad} = - \frac{2}{\lambda^2} \cos^2 \theta\, \sech^2 (x/\lambda) 
\end{equation}
for a Harris equilibrium. 
This agrees with the ideal MHD response, Eq.~(\ref{eq:vbs}), only in the limit of a standard symmetric layer $(x_s \rightarrow 0)$, i.e., the parallel mode. The cause of this disagreement is that the $k_\parallel \ll k$ assumption holds only near a resonant surface. This leads to substantial errors in the outer region, which also affects the dispersion relation. 

To illustrate that the primary difference between Galeev et al and our approach is the outer region solution, we first apply the $i \delta E_\parallel \simeq - \omega \delta \hat{A}/c$ approximation from Eq.~(\ref{eq:epaprx}), which was validated in the last section, to Eq.~(\ref{eq:agaleev}). This gives
\begin{equation}
(\partial_x^2 - k^2 - V_\textrm{ad}) \delta \hat{A} = - \delta \hat{A} \sum_s \frac{2 \omega_{ps}^2}{c^2} \frac{(\omega - \vc{k} \cdot \vc{U}_s)}{k_\parallel v_{Ts}} \zeta_s Z_1 (\zeta_s)  \label{eq:ahatgal}
\end{equation}
which is the same as the $U_{s,\parallel} \ll \omega/k_\parallel v_{Ts}$ limit of Eq.~(\ref{eq:dahat}) except in the outer region (left side) where $V_\textrm{ad}$ replaces $V_{bs}$.

Figure~\ref{fg:dacomp}a show solutions of the outer region of the Galeev equation 
\begin{equation}
\delta \hat{A}^{\prime \prime}  - (k^2 + V_\textrm{ad}) \delta \hat{A} = 0 \label{eq:gal_outer}
\end{equation} 
in comparison to ideal MHD from Eq.~(\ref{eq:fkr_outer}). For the parallel mode, the solutions are identical. However, the solutions are dramatically different for finite $\theta$, showing the inconsistency of the Galeev outer region equation with ideal MHD. To further emphasize this point, Fig.~\ref{fg:dacomp}b shows a comparison between $\Delta^\prime$ obtained from the numerical solutions of Eq.~(\ref{eq:gal_outer}), (\ref{eq:fkr_outer}), and the model solutions from Eq.~(\ref{eq:dph}) and Galeev's approximate solution \cite{foot:dpg}
\begin{equation}
\Delta^\prime_G \simeq - \frac{2 (k\lambda + \nu)}{\lambda} \frac{\Gamma [(k \lambda + \nu)/2]\, \Gamma [(1 + k\lambda - \nu)/2]}{\Gamma [ (k\lambda - \nu)/2]\, \Gamma [(1 + k\lambda + \nu)/2]}  \label{eq:dpg}
\end{equation}
where 
\begin{equation}
\nu \equiv \frac{1}{2} \bigl( \sqrt{1 + 8 \cos^2 \theta} - 1 \bigr) \label{eq:nu}  .
\end{equation}
This figure emphasizes that the $\Delta^\prime$ obtained from Eq.~(\ref{eq:gal_outer}) differs from the ideal MHD solution. It also shows that the approximate solution from Eq.~(\ref{eq:dpg}) requires that $k\lambda$ be sufficiently large to accurately model the numerical solution. 

\subsection{Electrostatic effects\label{sec:coppi}}

Previous work has considered the influence of electrostatic effects, showing that the ion acoustic wave can couple with the tearing mode, causing stabilization.\cite{buss:78,copp:79,lee:80,hosh:87} 
Electrostatic effects were eliminated from our analysis in Sec.~\ref{sec:decoupled} based on an argument of scale separation between the inner tearing layer and outer ideal region. 
These theories consider cases where the inner layer can be sufficiently broad that this scale separation is not satisfied. 
Indeed, we will see in the next section that the inner layer does broaden substantially for oblique modes, calling into question the neglect of electrostatic effects. 
It is therefore prudent to compare with these theories. 

\begin{figure}
\includegraphics[width=8.5cm]{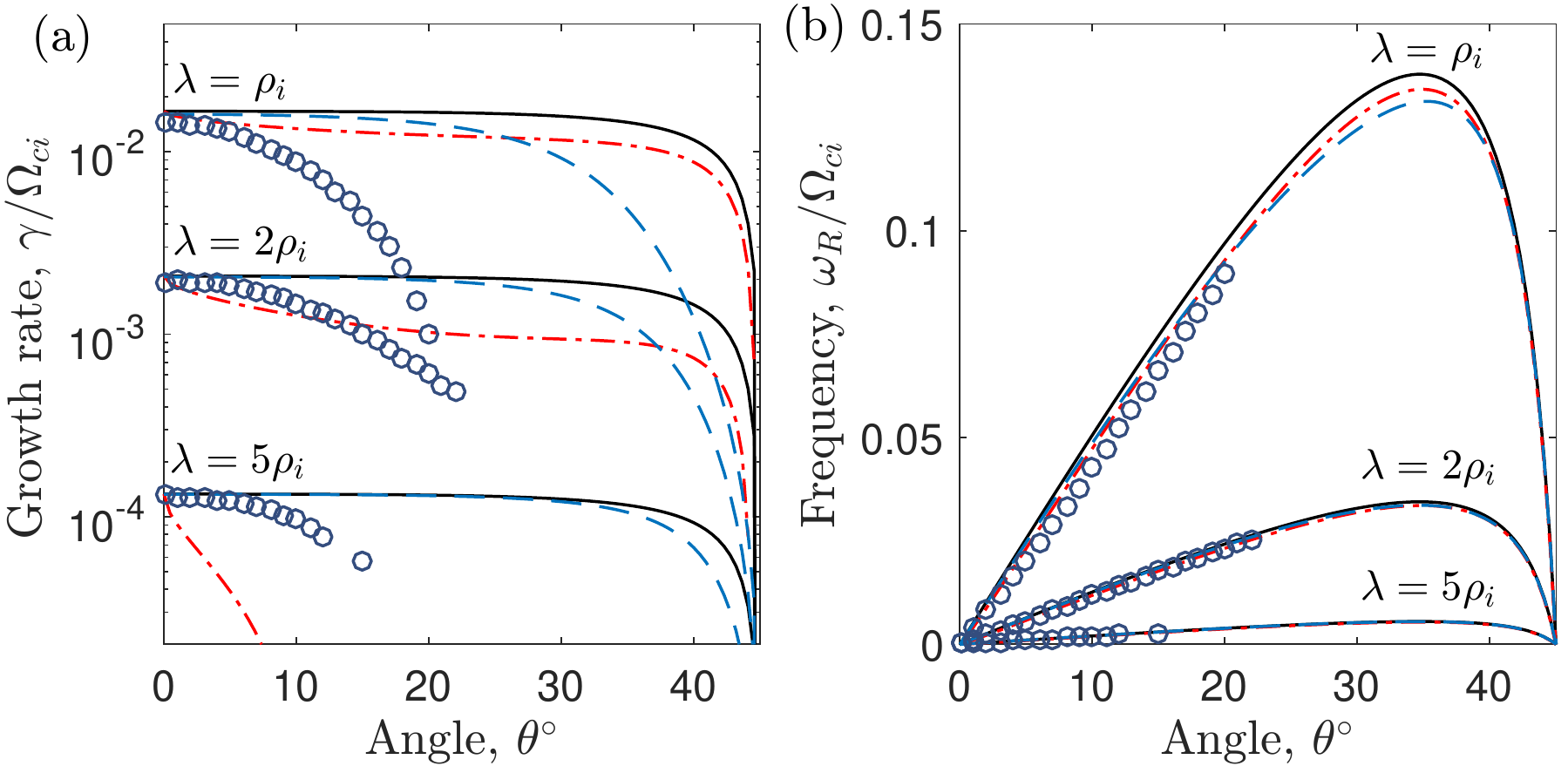}
\caption{(a) Growth rate and (b) real frequency computed from the VMID model (circles), DL theory from Eq.~(\ref{eq:dl_disp}) (black lines), the theory with electrostic effects from Eq.~(\ref{eq:lee}) (red dash-dotted lines) and the large $\Delta^\prime$ theory from Eq.~(\ref{eq:maha}) (blue dashed lines). Other parameters are: $m_i/m_e=1836$, $T_i/T_e = 1$, $k\lambda = 0.4$, $n_b = 0.3n_o$, and $B_{oy}/\bar{B}_{oz} = 1$.  }
\label{fg:disp_ml}
\end{figure}

Figure~\ref{fg:disp_ml} shows a comparison of the dispersion relation from Lee, Mahajan and Hazeline~\cite{lee:80}
\begin{equation}
\label{eq:lee}
\omega = \omega_{\textrm{DL}} - \sqrt{i} 0.85 \sqrt{|k_\parallel^\prime| c_s \rho_s} \sqrt{\frac{m_e}{m_i}} \frac{\omega_{o} + \vc{k} \cdot \vc{U}_i}{\sqrt{\omega_o}}
\end{equation}
where $k_\parallel^\prime = k/l_s$, $\rho_s = c_s/\Omega_{ci}$ and $\omega_o = \Re \lbrace \omega_{\textrm{DL}} \rbrace$. Here, $\omega_{\textrm{DL}}$ is the DL dispersion relation from Eq.~(\ref{eq:dl_disp}). The second term on the right side of Eq.~(\ref{eq:lee}) is the correction due to electrostatic effects. 
The figure shows that this theory does predict a smaller growth rate for oblique modes than DL. 
However, the predicted values do not agree well with the VMID solutions, or capture the trend of a sharp cutoff in the growth rate. 

Electrostatic effects do not appear to be the cause of stabilization of the oblique modes since the approximate theory neglecting electrostatic effects, Eq.~(\ref{eq:dahat}), captures the effect while boundary layer theories including electrostatics do not. 
Although the scale separation used to justify the neglect of electrostatic effects is similar to that used to justify boundary layer theory, it is not the same. 
Namely, in Eq.~(\ref{eq:epaprx}), the approximation made was $k_\parallel \delta \phi \ll \frac{\omega}{c} \frac{k_\eta}{k} \delta \hat{A}$. 
Although wide tearing layers at oblique angles access a range of values far from the current sheet, so $k_\parallel/k$ may not be small, the real frequency of oblique modes also increases approximately as $k_y U_e$. 
Thus, these effects compete in such a way that the electrostatic terms may remain small in comparison to the vector potential terms. 
Indeed, the electrostatic component of the eigenfunctions from the VMID solutions have been shown to be small in comparison to the vector potential contributions even at oblique angles (see Fig. 8 of Ref.~\onlinecite{daug:05}). 
This reference also showed that the ratio of the electrostatic to electromagnetic contribution to the eigenfunction scales linearly with $\omega_{pe}/\omega_{ce}$. Thus, one can alter the magnitude of the electrostatic terms with this parameter. 
Nevertheless, the computed growth rate remained independent of this parameter.~\cite{daug:05} 

\subsection{Nonconstant-$\psi$ approximation}

The DL theory applies the constant-$\psi$ approximation to obtain an analytic dispersion relation. 
This works best if the inner layer is very thin because $\delta \hat{A}$ does not vary significantly over a sufficiently short distance. 
For thicker inner layers, the approximation begins to break down. 
Methods have been developed to extend the theory, typically by keeping a linear term in the expansion of $\delta \hat{A}$ near the resonant surface. 
These theories are called large $\Delta^\prime$, or nonconstant-$\psi$ approximations. 
It was observed in the MHD theory that a nonconstant-$\psi$ approximation was required to accurately model oblique modes.~\cite{baal:12}
The next section will show that the inner layer of the collisionless problem becomes thicker at oblique angles. 
It is thus natural to investigate if nonconstant-$\psi$ effects are responsible for the observed stabilization. 

A large $\Delta^\prime$ theory of kinetic collisionless tearing modes has been computed in Mahajan \emph{et al}\cite{maha:79}, which provides the dispersion relation 
\begin{equation}
\label{eq:maha}
(\omega - \omega_o) \biggl(1 - i \frac{\Delta^\prime \omega}{\pi k_\parallel^\prime v_A} \biggr) = i \gamma_\textrm{DL} .
\end{equation}
Here, $\gamma_{\textrm{DL}} = \Im \lbrace \omega_{\textrm{DL}} \rbrace$, and $v_A = \bar{B}_{oz}/\sqrt{4\pi \bar{n}_o m_i}$ is the Alfv\'{e}n speed. 
Figure~\ref{fg:disp_ml} shows a comparison of the results of Eq.~(\ref{eq:maha}) with the VMID solutions. The nonconstant-$\psi$ modification does reduce the growth rate substantially at oblique angles. 
However, it still predicts a much broader spectrum of unstable modes than either VMID or Eq.~(\ref{eq:dahat}). 
Thus, the large $\Delta^\prime$ extension of DL theory does not appear able to explain the observed stabilization. 

\section{Analysis of Boundary Layer Theory\label{sec:const_psi}}

\subsection{Gyroradius effects} 

\begin{figure*}
\includegraphics[width=12cm]{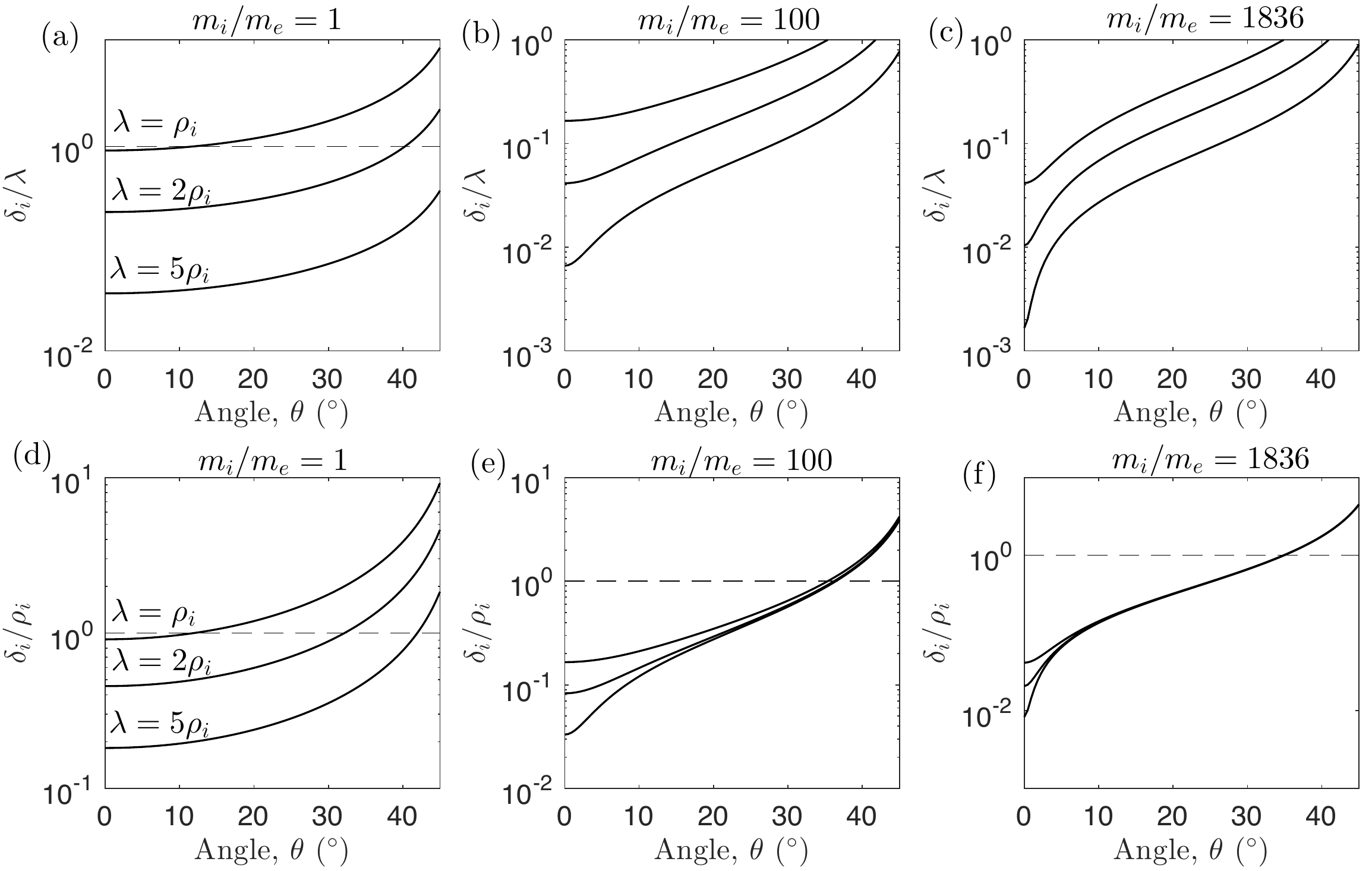}
\caption{Ion layer width computed from $\delta_i = |\omega| l_s/(k v_{Ti})$, compared with $\lambda$ (a)-(c) and $\rho_i$ (d)-(f) for each of the conditions shown in Fig.~\ref{fg:gam_gal}. In each case, the top line is for $\lambda = \rho_i$, the middle for $\lambda = 2 \rho_i$ and the bottom for $\lambda = 5 \rho_i$.}
\label{fg:delta_lambda}
\end{figure*}

A small gyroradius expansion was used in Sec.~\ref{sec:sgr} to reduce the full particle orbits from the Vlasov description in Eq.~(\ref{eq:s}) to Eq.~(\ref{eq:srho}). This is strictly valid only in the limit that both the electron and ion gyroradii are smaller than other length scales of relevance to the tearing mode. 
Figures~\ref{fg:delta_lambda}d-f show that this approximation is clearly violated at the conditions being explored in this paper. 
The largest gradient scale is the current sheet width, which for the parameters chosen in Fig.~\ref{fg:gam_gal} ranges from $\lambda = \rho_i$ to $5\rho_i$. Ions, then, are not strongly magnetized even on the scale of the shear of the current sheet. 
Electrons, however, are strongly magnetized at this scale for the large mass ratio cases since $\rho_e/\lambda \sim \sqrt{m_e/m_i} \ll 1$. However, use of the approximation at the scale of the inner reconnection layer faces a more stringent requirement. Figures~\ref{fg:delta_lambda}d-f plot an estimate of the ratio of the ion inner tearing layer width to the ion gyroradius, showing that in all cases considered the inner tearing layer width is smaller than the gyroradius. 
Since $\delta_i/\rho_i$ is independent of mass, the same result applies for electrons.
Here, the inner tearing layer width was estimated as in Drake and Lee, which provides $\delta_s \simeq |\omega| l_s/(kv_{Ts})$ where $\omega$ was computed from Eqs.~(\ref{eq:wdrake}) and (\ref{eq:gdrake}).\cite{drak:77} 
Although Fig.~\ref{fg:gam_gal} has shown that this does not accurately model the growth rate of oblique modes, it does accurately model the real mode frequency. Since the inner layer width estimate is dominated by the larger real component of the frequency, we expect that it remains a reasonable estimate at oblique angles.

Although the small gyroradius expansion is not expected to be valid at these conditions, comparison with the full VMID solutions in Fig.~\ref{fg:gam_gal} shows that the predictions resulting from it, namely Eq.~(\ref{eq:dahat}), capture the essential features of the growth rate profiles. 
The comparison reveals quantitative differences in the predicted growth rate of oblique modes, but it is quantitatively good for the parallel ($\theta=0^\circ$) mode (except at unity mass ratio), as well as the angle of obliquity at which stabilization occurs. 
The full orbit integrals are complicated, and any reduced analytic model must treat the orbits in an approximate fashion. The comparison in Fig.~\ref{fg:gam_gal} provides confidence that the small gyroradius expansion captures the essential features of interest from linear theory.

It may seem surprising that the growth rate derived from the small gyroradius expansion captures the essential features of the full VMID solution at conditions where the expansion is not expected to be valid. In this regard, one should notice that the linear growth rate depends on the total perturbed current within the resonance layer. Previous work has shown that the electron finite Larmour radius (FLR) broadens this layer, but does not significantly influence the total perturbed current.~\cite{kari:05}
Specifically, Fig.~1 of Ref.~\onlinecite{kari:05} shows that the perturbed current from the linear VMID calculation is smeared over a scale of a few $\rho_e$; it is not possible to form a narrower current channel. Although FLR effects do not seem to significantly influence the linear growth rate, they have bigger implications for the saturation. Exponential growth, as described by linear theory, is expected to saturate when the island width becomes comparable to the resonant layer thickness. Since FLR effects broaden the resonant layer, in comparison to the classical theoretical predictions, this allows modes to grow linearly to much larger amplitudes than would otherwise be expected. This issue is particularly relevant to magnetospheric layers, and has been discussed by Quest and Coroniti.~\cite{ques:81,ques:81b}

\subsection{Inner layer width}

Boundary layer theory is predicated on the assumption of an asymptotic scale separation between an ``outer region'' described by ideal MHD and an ``inner region'' where two-fluid and kinetic physics leads to violations of the frozen flux condition. 
Such a two-scale approximation was used not only in the theories of Sec.~\ref{sec:compare}, but also in Sec.~\ref{sec:sgr} which led to the decoupling of the electric and magnetic fields allowing one to describe the tearing mode from the single differential equation~(\ref{eq:dahat}). 
The possible absence of this scale separation is what led Coppi to suggest that a third electrostatic layer must be accounted for in some circumstances, as discussed in Sec.~\ref{sec:coppi}.

Figures~\ref{fg:delta_lambda}a-c show the ratio of the estimated inner tearing layer width ($\delta_i$) to the width of the current sheet ($\lambda$), which characterizes the scale of the ideal MHD outer region. 
At unity mass ratio, the scales are not well separated. 
We expect that this is the cause of the poor agreement between the VMID solutions and the solutions of Eq.~(\ref{eq:dahat}) at these conditions. 
For the large mass ratio cases $\delta_i/\lambda \ll 1$ near the center of the current sheet (small $\theta$), but the scale separation is violated at highly oblique angles. 
This calls into question the validity of this assumed scale separation at oblique angles.  
Despite the scale separation becoming questionable at oblique angles, the fact that Eq.~(\ref{eq:dahat}) captures the stabilization suggests that the influence of electrostatic terms is not the cause of this effect; see Sec.~\ref{sec:coppi}. 
Regarding the subsequent application of the boundary layer approximation in developing an analytic theory, specifically the DL theory from Sec.~\ref{sec:dl}, the approximation can be tested in more detail. 

To asses the boundary layer theories, we compare the left side of Eq.~(\ref{eq:dahat}), which has a characteristic scale of $\lambda$, with the inner region which is described by the right side 
\begin{equation}
\label{eq:Tx}
T(x) = - \lambda^2 \sum_s \frac{2\omega_{ps}^2}{c^2} \frac{(\omega - \vc{k} \cdot \vc{U}_s)}{k_\parallel v_{Ts}} \zeta_s Z_1(\zeta_s).
\end{equation}
(Here, the version of Eq.~(\ref{eq:dahat}) with $w_s$ replaced by $\zeta_s$ in the $Z$-function is used to make connection with the DL theory). 
In boundary layer theory, the spatially-dependent terms are expanded locally about the resonant surface: $k_\parallel \simeq k(x-x_s)/l_s$ and $\omega_{ps} \simeq \omega_{ps}(x_s)$, leading to an approximate expression describing the inner region 
\begin{equation}
\label{eq:Tdl}
T_{\textrm{DL}} (x) = -\lambda^2 \sum_s \frac{2 \omega_{ps}^2 |_{x_s} \omega (\omega-\vc{k} \cdot \vc{U}_s)}{c^2 k^2 v_{Ts}^2 (x-x_s)/l_s} Z_1 \biggl[ \frac{\omega l_s}{kv_{Ts}(x-x_s)} \biggr]
\end{equation}

\begin{figure}
\includegraphics[width=8.5cm]{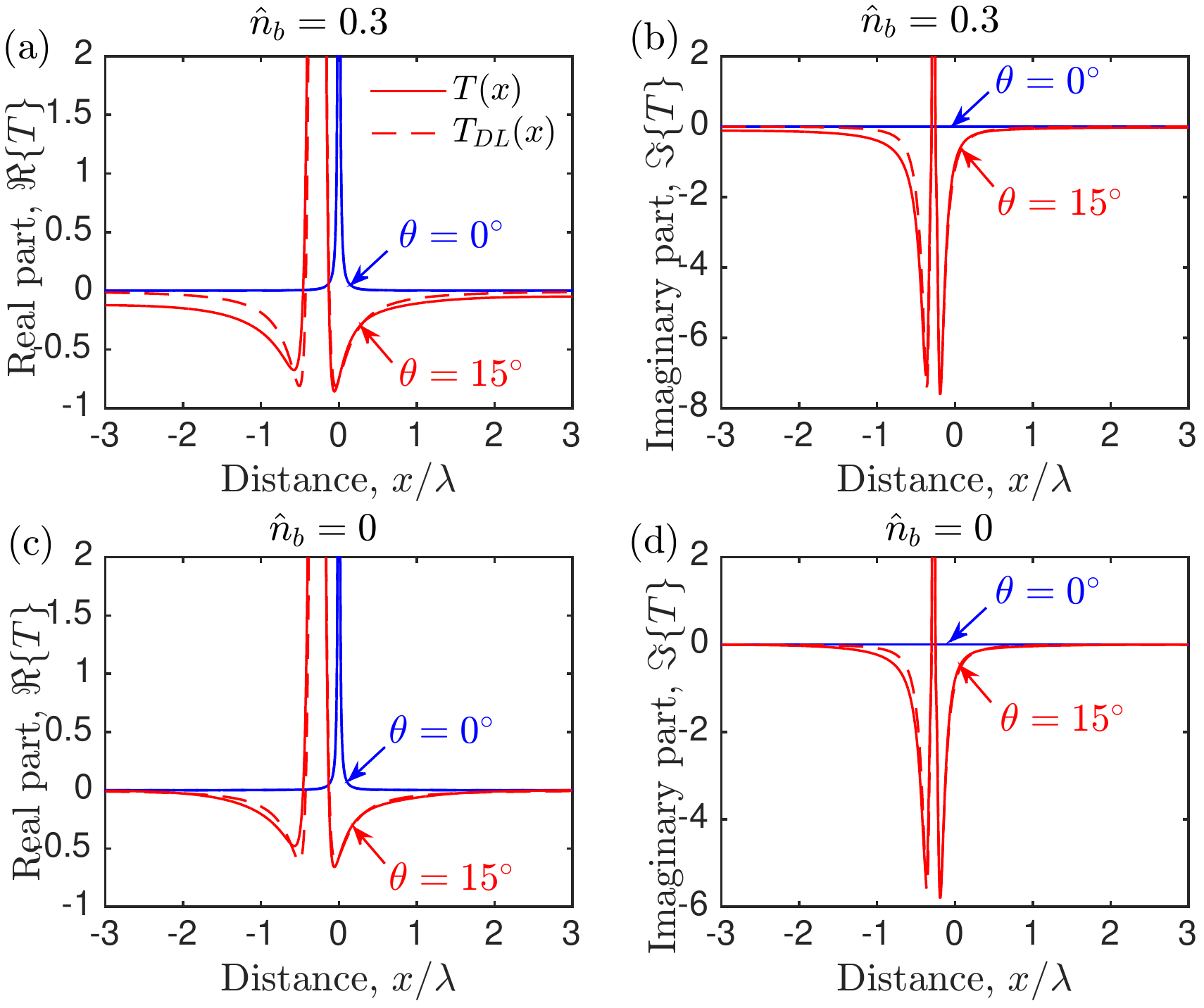}
\caption{(a) Real and (b) imaginary parts of $T(x)$ computed from Eq.~(\ref{eq:Tx}) (solid lines) and Eq.~(\ref{eq:Tdl}) (dashed lines) at $\theta = 0^\circ$ (blue) and $\theta=15^\circ$ (red) with parameters: $T_i/T_e = 1$,  $k\lambda = 0.4$, $B_{oy}/\bar{B}_{oz} = 1$, $\lambda = 2\rho_i$, and $\hat{n}_b = 0.3$. (c) and (d) show the same with $\hat{n}_b = 0$.  Dashed and solid lines are indistinguishable for $\theta=0^\circ$.}
\label{fg:bl}
\end{figure}

Figure~\ref{fg:bl}a-b shows a comparison of the solutions of Eqs.~(\ref{eq:Tx}) and (\ref{eq:Tdl}) for the same parameters as the $\lambda = 2\rho_i$ case of Fig.~\ref{fg:gam_gal}c. Here, the mode frequency is computed using the DL solutions from Eqs.~(\ref{eq:wdrake}) and (\ref{eq:gdrake}). 
This reiterates the breakdown of a scale separation at oblique angles, which was estimated in Fig.~\ref{fg:delta_lambda}. 
As $\theta$ approaches several degrees, near the value where the growth rate rapidly diminishes, the scale associated with the inner region $[T(x)]$ is observed to grow to a comparable value with the scale associated with the outer region ($\lambda$). 
This explains why boundary layer theory fails to accurately model the numerical solutions of Eq.~(\ref{eq:dahat}) at sufficiently oblique angles. 
The numerical solutions reveal that the growth rate rapidly diminishes at the same conditions; see Fig.~\ref{fg:gam_gal}. 

\subsection{Phase and the outer region~\label{sec:outer}} 

Figure~\ref{fg:bl} answers another inconsistency between the boundary layer and numerical solutions: Why is there a prominent complex (phase) component of $\delta \hat{A}$ for oblique modes ($\theta \neq 0$) in the numerical solutions, but this does not arise in the boundary layer theory? 

First, we observe that $T(x)$ is purely real for the parallel mode ($\theta = 0^\circ$) (as a result of $\omega$ being purely imaginary). Thus, no phase component is expected in this case, nor is one observed in the numerical solutions; see Fig.~\ref{fg:phase}. The figure also shows that $T(x=\pm \infty) \rightarrow 0$ and that $T(x)$ is localized if $\theta=0$. 
This case is amenable to a boundary layer analysis because the inner region decays to zero before reaching the scale of the outer region, and the inner and outer regions are both purely real and can be matched asymptotically. 
The boundary layer theory agrees with the numerical solutions in this case. 

Oblique modes have fundamentally different properties. For any finite angle, $T(x)$ becomes complex (has both real and imaginary components) implying that $\delta \hat{A}$ will be complex (i.e., have a finite phase). This is consistent with what is observed in the numerical solutions at finite angle; see Fig.~\ref{fg:phase}. 

\begin{figure}
\includegraphics[width=8.5cm]{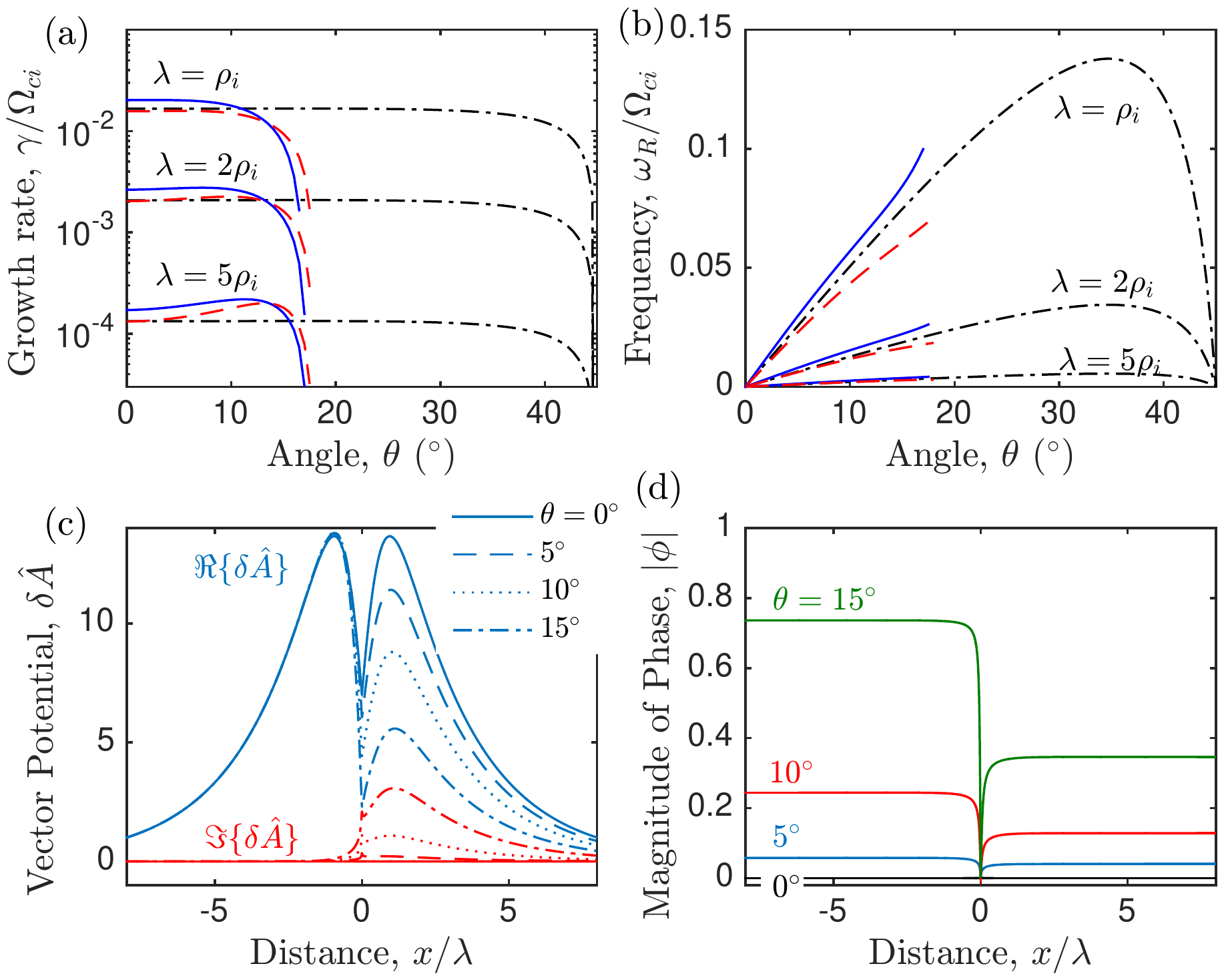}
\caption{(a) Growth rate and (b) real frequency computed from the decoupled layer equation~(\ref{eq:dahat}) (red dashed line), and the DL theory from Eq.~(\ref{eq:gdrake}) (black dash-dotted line) with parameters: $m_i/m_e = 1836$, $T_i/T_e = 1$,  $k\lambda = 0.4$, $n_b = 0.3n_o$, $B_{oy}/\bar{B}_{oz} = 1$. The blue solid lines show solutions with $n_b=0$. (c) shows the vector potential and (d) the phase magnitude from solving Eq.~(\ref{eq:dahat}) with $n_b=0$ and $\lambda = 5\rho_i$.}
\label{fg:nb0}
\end{figure}

In addition, Fig.~\ref{fg:bl} shows that at finite angles $T(x=- \infty) \neq 0$ if $\hat{n}_b \neq 0$. Since part of $T(x)$ extends far from the resonant surfaces, this implies that at finite angles the kinetic theory does not asymptote to the ideal MHD outer region. Instead, one should move the components of $T(x)$ extending to asymptotically far distances $T(x\pm \infty)$ to the left side of Eq.~(\ref{eq:dahat}) (the outer region). 
In fact, if one does not do this, the finite value of $T(x=- \infty)$ implies that the integral $\int dx T(x)$, which is used in the constant-$\psi$ approximation, actually diverges. 
This divergence happens to not arise in Drake and Lee's theory when integrating $\int dx T_{\textrm{DL}}(x)$ as a fortuitous consequence of the local expansion of $\omega_{ps}$ and $k_\parallel$ (which is not valid in this case). The local expansion forces $T_{\textrm{DL}} (x =\pm \infty) \rightarrow 0$ even at finite $\hat{n}_b$; see Fig.~\ref{fg:bl}a-b. 

\subsection{Role of background density}

Since $T(x)$ extends to asymptotically far distances only when $n_b \neq 0$, it is interesting to compare with the situation of no background density $n_b = 0$. 
This allows one to isolate the role of the imaginary component of $\delta \hat{A}$ to determine if it influences the growth rate, and in particular if it is associated with the damping observed at oblique angles. 
If $T(x)$ is sufficiently local near the resonant surface, one expects that  $\Im \lbrace \delta \hat{A} \rbrace$ will be small compared to the real component. 
Figure~\ref{fg:bl}c-d shows that $T(x)$ and $T_{\textrm{DL}}(x)$ agree quite well in this case. 
Figure~\ref{fg:nb0}c-d shows solutions of Eq~(\ref{eq:dahat}) with $\hat{n}_b=0$, confirming that $\Im \lbrace \delta \hat{A} \rbrace$ is small compared to the real component when $\theta$ is small (see also Fig.~\ref{fg:phase}), but that it grows comparable to the real component for $x>x_s$ at oblique angles. 
Figure~\ref{fg:nb0}a-b shows solutions of the dispersion relation computed from Eq~(\ref{eq:dahat}) with $\hat{n}_b=0$. 
These are found to damp at a similar angle as the case with $\hat{n}_b=0.3$ from Fig.~\ref{fg:gam_gal}c. 
This provides further evidence that the damping is caused by the broadening of the inner region, not simply the finite phase component.

\subsection{Role of guide field}  

\begin{figure}
\includegraphics[width=8.5cm]{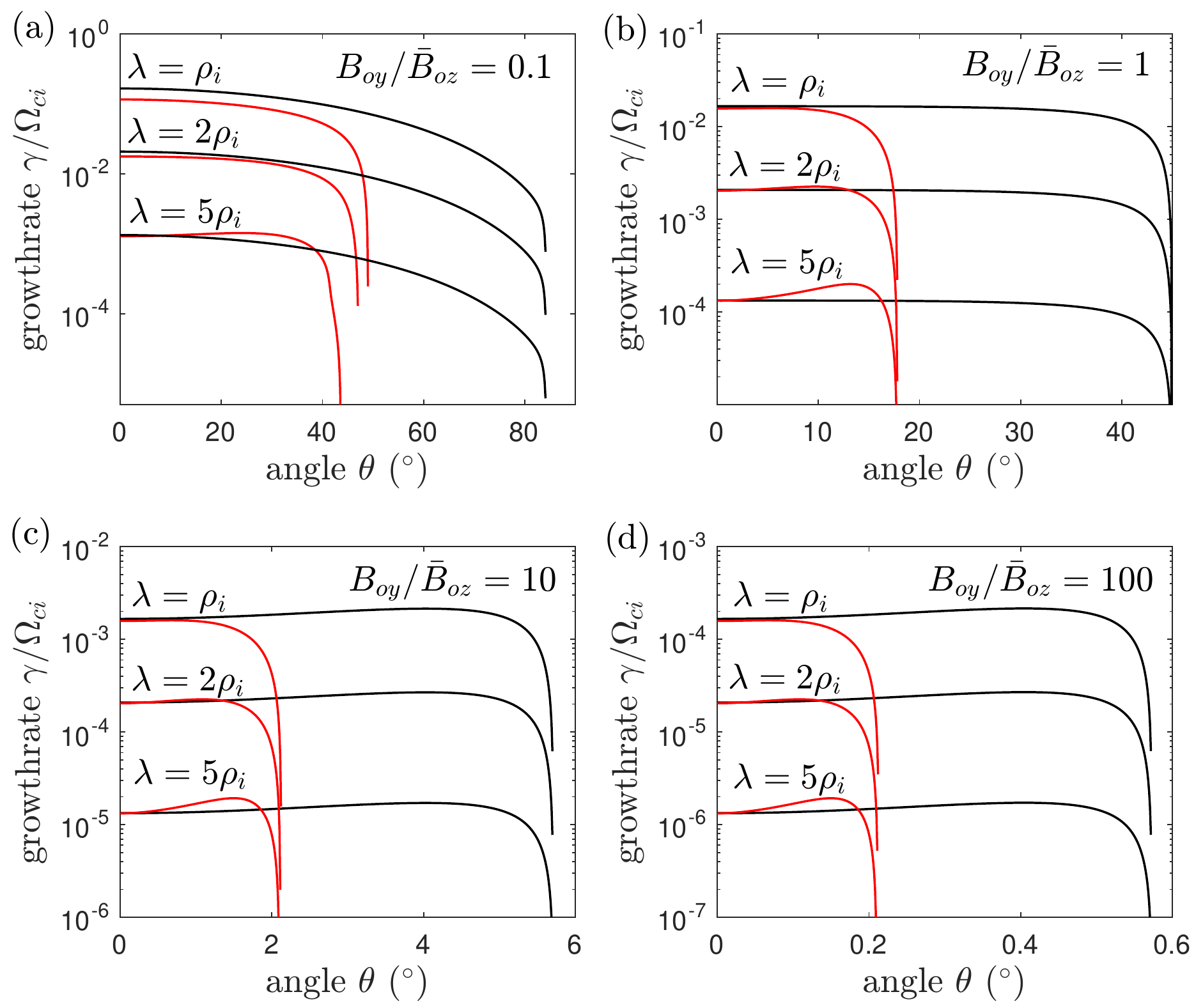}
\caption{Growth rate computed from Eq.~(\ref{eq:dahat}) (red lines) and the Drake and Lee theory from Eq.~(\ref{eq:gdrake}) (black lines) using $m_i/m_e = 1836$, $T_i/T_e = 1$, $k\lambda = 0.4$ and $n_b = 0.3n_o$ and four values of the guide field to asymptotic field ratio: (a) $B_{oy}/\bar{B}_{oz} = 0.1$, (b) 1, (c) 10 and (d) 100.}
\label{fg:guide}
\end{figure}

\begin{figure}
\includegraphics[width=8.5cm]{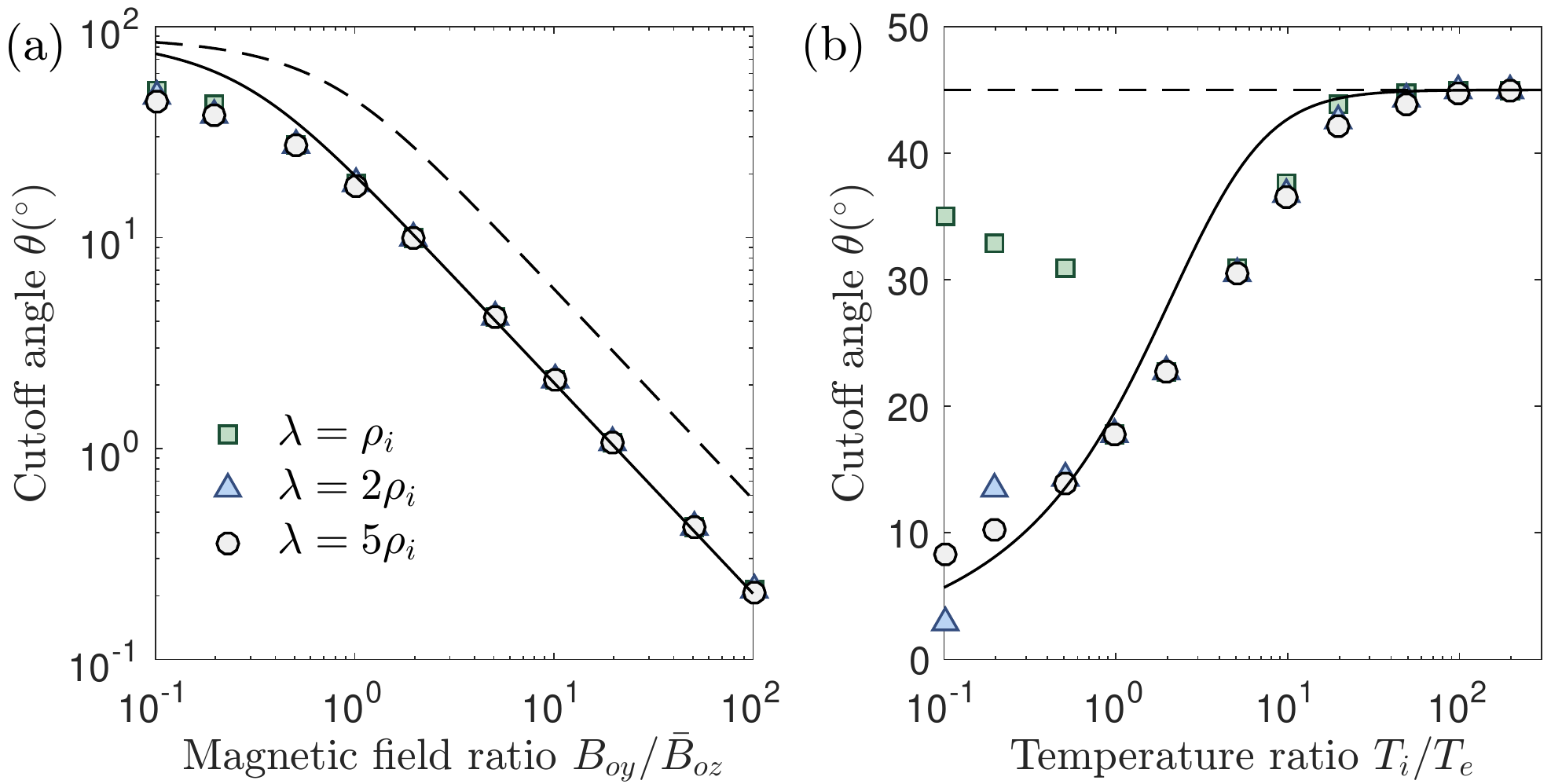}
\caption{ Cuttoff angle as a function of (a) $B_{oy}/\bar{B}_{oz}$ for $T_i/T_e = 1$, and (b) $T_i/T_e$ for $B_{oy}/\bar{B}_{oz} =1$. Other parameters include $m_i/m_e = 1836$, $n_b = 0.3 n_o$, and $k\lambda = 0.4$. Data points show solutions from Eq.~(\ref{eq:dahat}), the Drake and Lee cuttoff angle from Eq.~(\ref{eq:tc_dl}) (dashed lines) and from Eq.~(\ref{eq:mu_c}) (solid lines). }
\label{fg:theta_c}
\end{figure}

Figure~\ref{fg:guide} shows that the narrower spectrum of unstable modes predicted by Eq.~(\ref{eq:dahat}), in comparison to DL, persists over a wide range of guide field strength. 
For this range of parameters the cutoff angle is also independent of $\lambda/\rho_i$. 
Each of these observations is consistent with the notion that stabilization is associated with the width of the inner layer exceeding the shear scale $\lambda$. This is further quantified in Sec.~\ref{sec:threshold}. 
As the ratio of guide field to asymptotic field strength, $B_{oy}/\bar{B}_{oz}$, exceeds unity, the range of unstable angles narrows substantially simply because the field is predominantly in the guide field direction. This behavior is also shown in Fig.~\ref{fg:theta_c}, which displays the cutoff angle (where $\gamma = 0$) as a function of $B_{oy}/\bar{B}_{oz}$ from each theory.  

\subsection{Role of temperature ratio} 

\begin{figure}
\includegraphics[width=8.5cm]{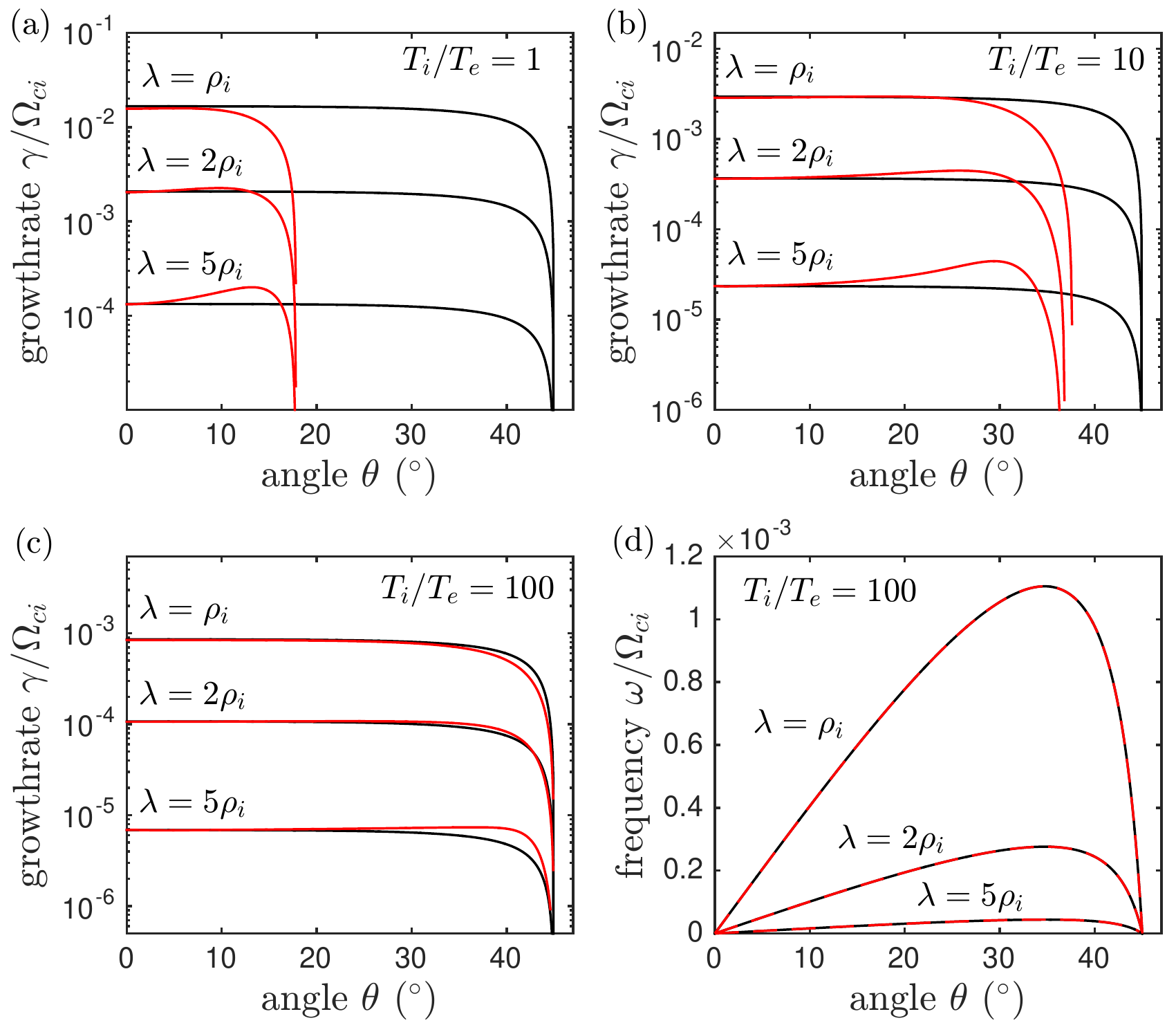}
\caption{Dispersion relation computed from Eq.~(\ref{eq:dahat}) (red lines) and the Drake and Lee theory from Eq.~(\ref{eq:gdrake}) (black lines) using $m_i/m_e = 1836$, $\bar{B}_{oz}/B_{oy} = 1$, $k\lambda = 0.4$ and $n_b = 0.3n_o$. (a) Growth rate for $T_i/T_e = 1$, (b) growth rate for $T_i/Te = 10$, (c) growth rate for $T_i/T_e = 100$. (d) Shows the real frequency for $T_i/T_e = 100$, where the results of the two theories are indistinguishable. }
\label{fg:temp}
\end{figure}

Figure~\ref{fg:temp} shows that the cutoff angle computed from Eq.~(\ref{eq:dahat}) depends significantly on the temperature ratio $T_i/T_e$. In contrast, the cutoff angle is independent of the temperature ratio in the DL model. At $T_i/T_e \sim 1$, the spectrum of unstable modes from Eq.~(\ref{eq:dahat}) is significantly more narrow than from the DL. However, as $T_i/T_e$ increases the spectrum predicted from Eq.~(\ref{eq:dahat}) broadens substantially. At a large temperature ratio ($T_i/T_e = 100$) the two results agree very well; see Fig.~\ref{fg:temp}c and d. The same trend is captured in Fig.~\ref{fg:theta_c}b, which shows the cutoff angle as a function of temperature ratio for three values of the ratio $\lambda/\rho_i$. The Drake and Lee limit is obtained at large temperature ratio. 
Since the scale separation between the inner region and the $\lambda$ scale is broader at large $T_i/T_e$, this also provides further evidence associating the cutoff in the growth rate with the broadening of the inner region. This will be further quantified in Sec.~\ref{sec:threshold}. 
This temperature ratio dependence is directly related to the drift stabilization. 
For a Harris sheet, the temperature ratio sets the ratio of flow velocities $U_i/U_e = - T_i/T_e$, while the ion velocity is fixed by the current sheet thickness $U_i/v_{Ti} = \rho_i/\lambda$. 
Thus, changing the temperature ratio effectively changes the electron drift speed $U_e = -U_i T_e/T_i$. 
For $T_i/T_e \gg 1$, $U_e$ is very small, and one returns the results of the DL theory valid in the limit $U_e \rightarrow 0$. 
Finally, Figs.~\ref{fg:theta_c}b and \ref{fg:temp} show that the cutoff angle is essentially independent of $\lambda/\rho_i$, when $T_i>T_e$, but that a significant dependence on this ratio arises when $T_i<T_e$. 

\subsection{Estimate of instability threshold\label{sec:threshold}} 

An important contribution that linear tearing theory can provide to the overall understanding of reconnection dynamics is the volume of space susceptible to tearing (or plasmoid formation), as this has significant consequences for subsequent heating processes as well as the volume susceptible to the formation of turbulence. This volume is determined by the cutoff angle ($\theta_c$) since the resonant surface furthest from the center of the current sheet that will be unstable is given by $x_c = \lambda \arctanh (\mu_c)$ where $\mu_c = \tan \theta_c B_{oy}/\bar{B}_{oz}$.  
We have shown that the differential equation~(\ref{eq:dahat}) provides a good description of the cutoff in comparison to the complete linear VMID solution. 
However, this requires solving the differential equation numerically. 
Here, we show that a simple analytic formula for $\theta_c$ can be obtained based on the results described previously in this section, which have established that damping occurs when the inner tearing layer becomes as wide as the current sheet ($\lambda$). 

The width of the inner layer ($\lambda^\prime$) can be estimated using Eq.~(\ref{eq:Tdl}). 
We approximate the stability condition as $|T(x=\lambda^\prime)| = c$, where $c$ is a constant that will be determined from a fit to the numerical solutions. 
Since $T\propto \lambda/|x-x_s|$, the numerical value of the cutoff width can be absorbed into the fit parameter. 
The outer layer of this region is determined by the condition that the argument of $Z_1$ becomes small, in which case $Z_1 \rightarrow 1$; see Fig.~\ref{fg:bl}. 
Applying the approximation that $\omega \simeq \omega_R \simeq k_yU_e$ at oblique angles and that $|x-x_s|/\lambda \simeq 1-\arctanh(\mu) \simeq 1 - \mu$, the cutoff condition is then 
\begin{equation}
\label{eq:c1}
T \simeq \frac{4\mu^2}{1-\mu} \biggl( \frac{1-\mu^2 + \hat{n}_b}{1-\mu^2} \biggr) \frac{T_e/T_i}{1+T_i/T_e} = c .
\end{equation}
Finally a simplification can be made based on the previous observation that the background density does not significantly influence the cutoff angle (see Fig.~\ref{fg:nb0}), so the term in parenthesis can be approximated as 1. Applying this, Eq.~(\ref{eq:c1}) provides the following estimate for the cutoff angle
\begin{equation}
\label{eq:mu_c}
\mu_c = \frac{c}{8} \frac{T_i}{T_e} \biggl(1 + \frac{T_i}{T_e} \biggr) \biggl[ \sqrt{1 + \frac{16}{c} \frac{T_e/T_i}{(1+T_i/T_e)}} -1 \biggr]
\end{equation}
such that instability is expected when $\theta \lesssim \theta_c = \arctan (\mu_c \bar{B}_{oz}/B_{oy})$. 
Alternatively, the critical temperature ratio at a given angle and layer thickness ($\lambda/\rho_i$) can be related to a critical electron drift speed since $U_e = - \frac{T_e}{T_i} U_i = - \frac{T_e}{T_i} \frac{\rho_i}{\lambda}$. 

Figure~\ref{fg:theta_c} shows a comparison between the cutoff angle predicted by Eq.~(\ref{eq:mu_c}) and the numerical results of Eq.~(\ref{eq:dahat}). The fit parameter used was $c=0.4$, which is a reasonable number based on the fact that $T(x) = c$ defines where the effective with of $T$ is measured (see Fig.~\ref{fg:bl}). Also shown in the figures is the cutoff from the DL dispersion relation, which is simply $\mu = 1$, leading to 
\begin{equation}
\label{eq:tc_dl}
\theta_{c,\textrm{DL}} = \arctan(\bar{B}_{oz}/B_{oy})  .
\end{equation} 
Figure~\ref{fg:theta_c}a shows that Eq.~(\ref{eq:mu_c}) captures both the large $B_{oy}/\bar{B}_{oz}$ scaling and the value at which the deviation from this scaling occurs (for $B_{oy}/\bar{B}_{oz} < 1$). Importantly, Fig.~\ref{fg:theta_c}b shows that Eq.~(\ref{eq:mu_c}) accurately captures the temperature ratio dependence of the cutoff angle (when $T_i/T_e$ is sufficiently large).
Equation~(\ref{eq:c1}) shows that the thickness of the inner region scales as $1/(T_i/T_e)^2$ when $T_i/T_e \gg 1$. 
That is, the inner region becomes very thin at large temperature ratio, and one should expect in this limit that the boundary layer scale separation that the analytic dispersion relations (such as DL) are based on will be valid. 
Indeed, this is what is observed, as the two predictions merge at high temperature ratio; see Fig.~\ref{fg:theta_c}b. 
For $T_i/T_e \ll 1$, a dependence on the layer thickness is observed that is not captured by Eq.~(\ref{eq:mu_c}).
This figure provides further evidence that broadening of the tearing layer, associated with the diamagnetic drift, is responsible for the stabilization at oblique angles. 
It also shows that Eq.~(\ref{eq:mu_c}) provides a reliable estimate for the cutoff angle of oblique collisionless tearing modes. 

\section{Conclusions} 

It was shown that oblique collisionless tearing instabilities of a Harris current sheet are stabilized due to the density-gradient-driven diamagnetic drift. 
The complicated Vlasov-Maxwell system of equations was simplified to a single second order differential equation for the vector potential perturbation, which captures the essential features of the dispersion relation obtained from the full numerical solution of the linear Vlasov-Maxwell equations. 
Theories from previous literature are not able to capture this stabilization because they are based on a boundary layer scale separation between an inner tearing layer and an outer ideal MHD region. 
The inner layer was observed to broaden at oblique angles, invalidating this approximation. 

These results are applicable to studies of magnetic reconnection in space and fusion plasmas, where tearing and plasmoid instabilities are known to initiate and accelerate the rate of magnetic reconnection. 
They provide a method to compute the spectrum of oblique modes, which contributes to theories of turbulence generation and particle acceleration. 
They are particularly relevant to understanding the magnetopause, where reconnection occurs within thin asymmetric layers in which diamagnetic drifts are strong. 
In this context, Galeev coined the term ``percolation'' to denote the regime in which many overlapping islands formed by tearing modes give rise to turbulent particle transport.~\cite{gale:86} 
Our results show that diamagnetic drifts broaden the inner tearing layer, so that in current sheets with strong diamagnetic drifts only resonant surfaces with small $\vc{k} \cdot \vc{U}$ are expected to be unstable. 
This leads to a narrower spectrum of instabilities, i.e., a smaller volume of space susceptible to tearing, than was previously expected. 
A simple analytic estimate for the stabilization threshold was provided in Eq.~(\ref{eq:mu_c}), which can be used to predict the volume of space susceptible to linear tearing instabilities.  
The results also highlight the importance of the equilibrium current sheet properties in understanding tearing instabilities, and by extension the overall reconnection process. 
Specifically, drift-stabilization is not present in other common equilibria used to model such processes, such as the force-free current sheet.~\cite{liu:13}

Interesting questions remain to be answered, such as to what extent fluid theory can describe collisionless tearing and to what extent purely kinetic effects such as Landau damping play a role.  
For instance, Landau damping is a non-negligible, and even prominent, contributor to the the dispersion relation [arising in the plasma dispersion function of Eq.~(\ref{eq:dahat})] in the kinetic theory described in this work. 
Yet, two fluid theories have found success in modeling collisionless tearing modes in some geometries.~\cite{akca:16}
In one case the predicted growth rates differ only by a factor of $\sqrt{\pi}$.~\cite{wang:93} 
It would be useful to understand if two-fluid theory captures the stabilization of oblique modes observed in this work, and if that would lead to a simplified dispersion relation. 
It would also be important to revisit the linear tearing stability theory in the presence of temperature gradients. 
Although the Harris sheet does not have them, both the magnetopause and fusion machines have density and temperature gradients. 

\begin{acknowledgments}

This material is based upon work supported by the U.S. Department of Energy, Office of Science, Office of Fusion Energy Sciences under a Postdoctoral Research Program administered by the Oak Ridge Institute for Science and Education, DOE award number DE-SC0016159, and by NSF grant no. AGS-0962698. Contributions from W.~D.~were supported by the Basic Plasma Science Program from the DOE Office of Fusion Energy Sciences.

\end{acknowledgments}

\bibliography{refs.bib}

\begin{thebibliography}{99}
\bibitem{note1:gale} Ref.~\onlinecite{gale:86} takes $\hat{\vc{b}}^s$ to be the $\parallel$ direction in defining $\delta \hat{A}_\parallel$. However, the $k_\parallel$ that is cited must be parallel to  $\vc{B}_o$ since $\vc{k} \cdot \hat{\vc{b}}^s = 0$.
\bibitem{foot:dpg} The factor of 2 appearing in front of Eq.~(\ref{eq:dpg}) is missing in Ref.\ \onlinecite{gale:86} (see equation 14), but is required to get the $\theta = 0$ results correct.
\end{thebibliography}

\end{document}